\documentclass[
    article,
    reprint,
    twocolumn,
    aps,
    prb,
    superscriptaddress,
    amsmath,amssymb,
    longbibliography,
    ]{revtex4-2}
\usepackage{graphicx}
\usepackage{dcolumn}
\usepackage{siunitx}
\usepackage{float}
\usepackage{bm}
\usepackage{amsmath}
\usepackage{mathtools}
\usepackage{afterpage}
\usepackage[utf8]{inputenc}
\usepackage{bm}
\usepackage{soul}
\usepackage{color}
\usepackage{lipsum}
\usepackage{booktabs}
\usepackage{multirow}

\begin{document}

\title{Investigation of nonreciprocal spin-wave dynamics in Pt/Co/W/Co/Pt multilayers}

\author{Krzysztof~Szulc}
\email{krzysztof.szulc@amu.edu.pl}
\affiliation{%
Faculty of Physics, Adam Mickiewicz University, Pozna\'{n}, Uniwersytetu Pozna\'{n}skiego 2, 61-614 Pozna\'{n}, Poland 
}%
\author{Simon~Mendisch}
\affiliation{%
Department of Electrical and Computer Engineering, Technical University of Munich (TUM), Arcisstr. 21, 80333 Munich, Germany
}%
\author{Micha\l{}~Mruczkiewicz}
\affiliation{%
Institute of Electrical Engineering, Slovak Academy of Sciences, 841 04 Bratislava, Slovakia
}%
\affiliation{%
Centre for Advanced Materials Application (CEMEA), Slovak Academy of Sciences, Dúbravská cesta 5807/9, 845 11 Bratislava, Slovakia
}%
\author{Francesca~Casoli}
\affiliation{%
Istituto dei Materiali per l'Elettronica ed il Magnetismo (IMEM-CNR), I-43124 Parma, Italy
}%
\author{Markus~Becherer}
\affiliation{%
Department of Electrical and Computer Engineering, Technical University of Munich (TUM), Arcisstr. 21, 80333 Munich, Germany
}%
\author{Gianluca~Gubbiotti}
\affiliation{%
Istituto Officina dei Materiali del CNR (CNR-IOM), Sede Secondaria di Perugia, c/o Dipartimento di Fisica e Geologia, Universit\`{a} di Perugia, I-06123 Perugia, Italy
}

\begin{abstract}
We present a detailed study of the spin-wave dynamics in single Pt/Co/W and double Pt/Co/W/Co/Pt ferromagnetic layer systems. The dispersion of spin waves was measured by wavevector-resolved Brillouin light scattering spectroscopy while the in-plane and out-of-plane magnetization curves were measured by alternating gradient field magnetometry. The interfacial Dzyaloshinskii-Moriya interaction induced nonreciprocal dispersion relation was demonstrated for both single and double ferromagnetic layers and explicated by numerical simulations and theoretical formulas. The results indicate the crucial role of the order of layers deposition on the magnetic parameters. A significant difference between the perpendicular magnetic anisotropy constant in double ferromagnetic layer systems conduces to the decline of the interlayer interactions and different dispersion relations for the spin-wave modes. Our study provides a significant contribution to the realization of the multifunctional nonreciprocal magnonic devices based on ultrathin ferromagnetic/heavy-metal layer systems.
\end{abstract}

\maketitle

\section{Introduction}

High demand for improvement in the storage and computing devices and decrease of their power consumption leads to continued interest in spintronic phenomena. Recently, the interfacial Dzyaloshinskii–Moriya interaction (iDMI) \cite{Moriya60} brought the attention of the researchers as it permits to stabilize topological magnetic solitons, e.g., skyrmions and radial vortices \cite{Finocchio16}. The iDMI is an asymmetric exchange interaction and can be induced in ultrathin multilayer systems where the inversion symmetry is broken between both interfaces of the ferromagnetic layer. It is induced by large spin-orbit coupling between a ferromagnet and heavy-metal atoms at one of the interfaces of an ultrathin ferromagnet. The energy contribution of iDMI is minimized when spins are aligned perpendicularly in specific direction, described by the equation $E_{\rm{DMI}}=-\sum_{i,j} \mathbf{D}_{ij} \cdot (\mathbf{S}_{i} \times \mathbf{S}_{j})$, where $\mathbf{D}_{ij}$ is the Dzyaloshinskii–Moriya vector. Therefore, the orientation of the $\mathbf{D}_{ij}$ determines whether right-handed or left-handed rotation sense between neighboring spins is the configuration of lower energy. Even if the iDMI favors noncollinear alignment of spins when it is strong, the single-domain state can be achieved at high magnetic fields. Then, the chiral property of the iDMI is exhibited in nonreciprocal spin-wave dynamics \cite{santos2020nonreciprocity,udvardi2009,cortes-ortuno2013,mruczkiewicz2016} and can be used to tailor the magnon dispersion relation, and thus used to extend the functionality of the magnonic devices \cite{szulc2020spin}.

The iDMI value is one of the highest in Pt/Co systems, which are also characterized by strong perpendicular magnetic anisotropy (PMA). Thus, many of the iDMI systems are basing on Pt/Co ultrathin films with broken inversion symmetry, e.g., Pt/Co/AlO\textsubscript{x} \cite{belmeguenai2015,cho2015thickness}, Pt/Co/MgO \cite{boulle2016room} Pt/Co/Ta \cite{woo2016observation}, Pt/Co/(W,Ta,Pd) \cite{zhang2017}, Pt/Co-Ni/Ta \cite{yu2016spin}, Ir/Fe/Co/Pt \cite{soumyanarayanan2017tunable}, Pt/Co/Cu/AlO\textsubscript{x} \cite{kim2017}, Pt/Co/Os/Pt \cite{tolley2018}, Pt/Cu/Co/Pt \cite{parakkat2016}. In Ref.~[\onlinecite{Legrand2018}] the implications of asymmetric multilayers [Pt/Co/Ir]$_5$ with broken inversion symmetry on domain wall chirality, skyrmion stability, and its dynamics were reported. Interestingly, weak iDMI was also present in symmetric Pt/Co/Pt multilayers \cite{wells2017} due to different quality of the Co interfaces. The influence of the surface quality can also have a significant impact on the anisotropy of the sample \cite{danilov2018}.

Most of the studies, also cited above, considered structures with sandwich multilayers, where the unit Pt/Co/X was repeated $n$ times, and magnetic parameters of Co layers had negligible differences.  Nevertheless, the stack of layers with deliberately different magnetic properties of magnetic materials across the thickness might also have interesting properties especially related to the magnetization and spin-wave dynamics, useful for applications in spintronics and magnonics as indicated by recent studies. For example, in the antiferromagnetic-exchange-coupled symmetric bilayer CoFeB/Ru/CoFeB, the measured Ru thickness-dependent nonreciprocity was related to the difference between perpendicular interface anisotropy of the bottom and top CoFeB layer \cite{belmeguenai2017}. Thus, in this paper, we will investigate the spin-wave properties in systems composed of two ferromagnetic layers, i.e., double ferromagnetic layer (DFL) systems.

DFL systems with opposite sign of iDMI in the limit of noninteracting layers shall exhibit two nonreciprocal dispersion-relation branches, that are mirror images at wavevector $k=0$. Therefore, having a dispersion resembling the electronic Rashba splitting is expected \cite{santos2020modelling,costa2010,henry2019}. Additionally, the coupling strength between the layers can be tuned by Ruderman-Kittel-Kasuya-Yosida (RKKY) interaction or dipolar interaction \cite{grimsditch1996,fuchs1997}. Also, interlayer DMI interaction in the DFL system, competing with RKKY interaction, was recently shown \cite{fernandez2019symmetry}. These works point out that magnonic dispersion can be tailored in many ways in the DFLs and multilayers with varied iDMI, magnetic properties of the sublayers, and the coupling strength between them. 

The strength of iDMI can be quantified by several methods, such as domain-wall velocity \cite{karnad2018,shahbazi2019}, asymmetric hysteresis loop method \cite{han2016} and magnetic force microscopy \cite{samardak2018}. A detailed review of the different techniques used to measure iDMI in ultrathin films has been recently published by \citet{kuepferling2020measuring}. Among them, Brillouin light scattering (BLS) spectroscopy has demonstrated to be a powerful and reliable technique since it combines the high sensitivity to detect signals from spin waves in magnetic monolayers \cite{krams1992,madami2004,nembach2015linear,di2015,tacchi2017,kumar2020}, and the possibility to explore a wide range of wavevectors ($0-\SI{100}{\radian/\micro\meter}$) and frequencies ($1-500$ GHz). The estimation of iDMI constant from experimental measurements was studied also for complex multilayers \cite{bouloussa2018}.

In this work, we study Pt/Co/W/Co/Pt DFL systems and, as the reference samples, the Pt/Co/W single ferromagnetic layer (SFL) systems. Hysteresis loops were measured using the alternating gradient force magnetometer (AGFM). BLS is used to measure the dispersion relation of spin waves and to extract the iDMI constant by measuring the frequency asymmetry ($\Delta f$) between the spin waves propagating in opposite directions. The numerical simulation in the time and frequency domain provides an interpretation of the experimental results and allows us to extract effective magnetization and anisotropy constant of the Co layers. In particular, we found that in DFL samples, the Co layers have an opposite sign of iDMI due to the inversion of interfaces and significantly different anisotropies due to a different order of the layers deposition. With these properties, we aim to demonstrate systems with two, almost independent spin-wave dispersion relations possessing nonreciprocal properties.

\section{Experimental methods}

The characterized thin films were deposited at room temperature via confocal rf-magnetron sputtering (base pressure $< \SI{2e-7}{\milli\bar}$) onto $n^{-}$ doped silicon $(100)$ substrates with an in-house grown thermal oxide (thickness $\approx \SI{20}{\nano\meter}$). Before the deposition, residual water was removed from the samples using a $\SI{300}{\electronvolt}$ Ar$^{+}$ ion beam. All materials were deposited at a constant argon pressure $\SI{4}{\micro\bar}$ ($\approx \SI{3}{ \, \milli Torr}$) except for the Ta adhesive, which was deposited at $\SI{2}{\micro\bar}$ ($\approx \SI{1.5} {\, \milli Torr}$). The rf-power applied to the 2-inch targets was identical for all materials ($\SI{40}{\watt}$). To reduce contaminants the dead times between elements during the automated deposition were generally kept below $\SI{1}{\second}$. All multilayer stacks feature a standard adhesive ($ \SI{1.5}{\nano\meter}$ Ta), seed ($ \SI{6}{\nano\meter}$ Pt), and capping ($ \SI{3}{\nano\meter}$ Pt) layer. The stacks are, therefore, solely addressed by their magnetic layer compositions with thicknesses given in nanometers (e.g., Co(1.6)/W(0.95)/Co(1.95)). Single ferromagnetic Co layers having a thickness of 1.6 and \SI{1.95}{\nano\meter} have also been grown on Pt films and used as reference samples. The set of four fabricated samples is listed in Table~\ref{tab:params}.

Hysteresis loops were measured by the AGFM, applying a magnetic field up to 15 kOe in the directions parallel and perpendicular to the film plane.

\begin{figure}
    \centering
    \includegraphics[width=0.8\linewidth]{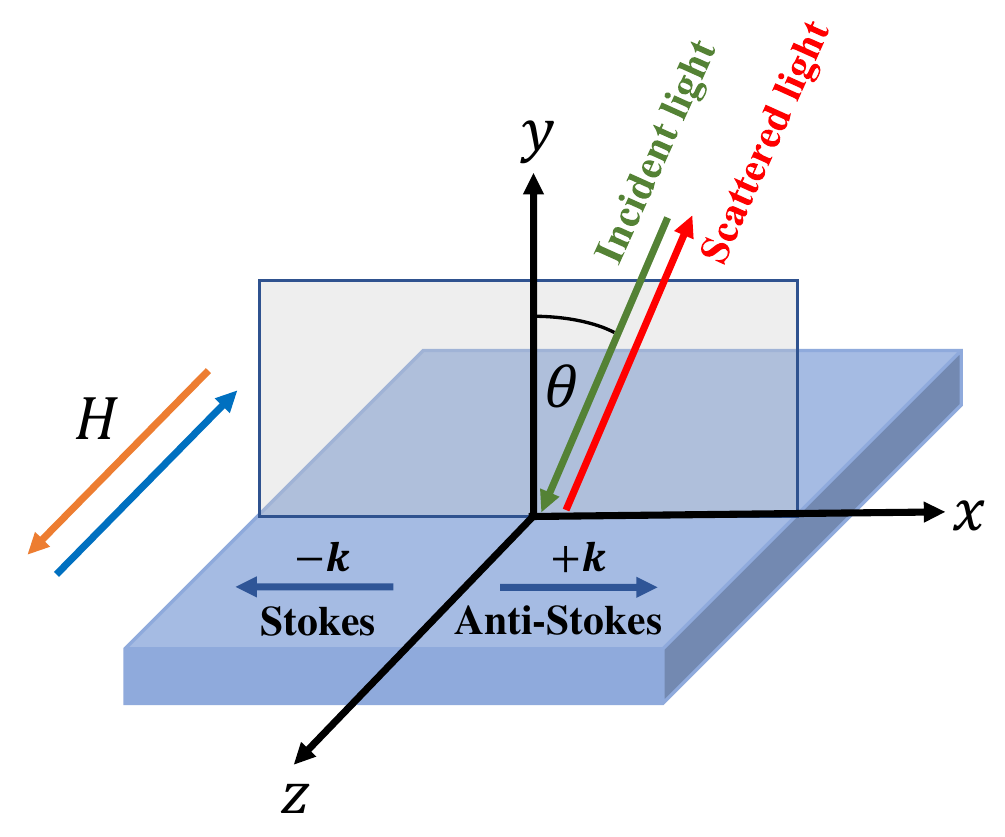}
    \caption{Schematic drawing of the BLS scattering geometry in the Damon-Eshbach configuration. The incident light makes an angle $\theta$ with respect to the sample normal. Measurements are performed in the backscattering configuration, where the same camera objective is used to focus laser light upon the sample surface and to collect scattered light sent to the interferometer for the frequency analysis. A magnetic field $H$ is applied in the sample plane and perpendicular to the incidence plane of light.}
    \label{fig:BLS_configuration}
\end{figure}

BLS spectra were recorded at room temperature in the backscattering configuration using a (3+3)-tandem Sandercock-type interferometer \cite{sandercock1982}. About 200-mW p-polarized monochromatic light from a solid-state laser $\lambda = \SI{532}{\nano\meter}$ was focused onto the sample surface. An in-plane magnetic field $H$ was applied parallel to the sample surface and perpendicular to the plane of incidence of light in the so-called Damon-Eshbach configuration. A schematic representation of the scattering geometry is represented in Fig.~\ref{fig:BLS_configuration}. Due to in-plane momentum conservation, the wavevector $k$ of spin waves entering into the scattering process is given by $k = (4\pi/\lambda) \sin \theta$. Spin waves traveling in the $-x$ and $+x$ directions appear as peaks in the Stokes and anti-Stokes side of the spectra, respectively. BLS measurements were performed in two different configurations: 1) by changing the magnitude of the external magnetic field applied in the sample plane at normal incidence of light upon the sample surface ($\theta = \SI{0}{\degree}$), i.e. $k=0$; 2) by sweeping the wavevector $k$ in the range from 0 to $\SI{20}{\radian/\micro\meter}$ at fixed applied field $H = \pm 5.5$ kOe \cite{Carlotti_2002}. Reversing the direction of the external magnetic field from $+5.5$ to $-5.5$ kOe is equivalent to reverse the direction of the propagating spin waves. The frequency asymmetry induced by iDMI is proportional to the sine function of the in-plane angle $\phi$ between the applied field and the wavevector direction \cite{cortes-ortuno2013}. In the Damon-Eshbach configuration ($\phi = \SI{90}{\degree}$), $\Delta f$ is maximum.

\section{Theoretical and numerical methods}

Spin-wave dynamics are calculated numerically using the finite-element method in COMSOL Multiphysics \cite{graczyk2018}. Motion of the spin is described by the Landau-Lifshitz equation:
\begin{equation}\label{eq:LL}
    \frac{\partial\textbf{M}}{\partial t} = -\gamma \mu_0 \textbf{M} \times \textbf{H}_{\text{eff}},
\end{equation}
where $\textbf{M} = (m_x,m_y,m_z)$ is the magnetization vector, $\gamma$ is the gyromagnetic ratio, $\mu_0$ is the magnetic permeability of vacuum, and $H_{\text{eff}}$ is the effective magnetic field, which is given as follows:
\begin{equation}
\begin{split}
    \textbf{H}_{\text{eff}} = H \mathbf{\hat{z}} + \frac{2A_{\text{ex}}}{\mu_0 M_{\text{S}}^{2}}\nabla^2\textbf{M} + \frac{2D}{\mu_0 M_{\text{S}}^{2}} \, \mathbf{\hat{z}} \times \frac{\partial \textbf{M}}{\partial x} + \\ + \frac{2K_u}{\mu_0 M_{\text{S}}^{2}} m_y \mathbf{\hat{y}} - \nabla \varphi,
\end{split}
\end{equation}
where $H$ is the external magnetic field, $M_{\text{S}}$ is the saturation magnetization, $A_{\text{ex}}$ is the exchange stiffness constant, $D$ is the iDMI constant, $K_u$ is the PMA constant, and $\varphi$ is the magnetic scalar potential fulfilling Poisson-like equation
\begin{equation}\label{eq:pot}
    \nabla^2 \varphi = \nabla\cdot\textbf{M}.
\end{equation}
In the DFL structures, the RKKY interaction is applied as the boundary conditions on the inner interfaces of the ferromagnetic layers \cite{klingler2018} along with the boundary condition for the exchange interaction
\begin{equation}
\begin{split}
    0 = 2 A_{\text{ex}} \mathbf{\hat{z}} \times \frac{\partial \mathbf{M}_{1(2)} (y)}{\partial y} \bigg|_{y=y_{1(2)}^{\text{in}}} - \\ - J \mathbf{\hat{z}} \times (\mathbf{M}_{2(1)} - \mathbf{M}_{1(2)}(y_{1(2)}^{\text{in}})),
\end{split}
\end{equation}
where $J$ is the RKKY constant and subscripts numerate bottom and top layers in DFL samples.
The boundary condition on the outer interfaces consist only of the exchange interaction term
\begin{equation}
    0 = 2 A_{\text{ex}} \mathbf{\hat{z}} \times \frac{\partial \mathbf{M}_{1(2)} (y)}{\partial y} \bigg|_{y=y_{1(2)}^{\text{out}}}
\end{equation}
\\
Eqs.~(\ref{eq:LL}) and (\ref{eq:pot}) are solved in the linear approximation, i.e., assuming $m_x,m_y \ll m_z \approx M_S$. Time-domain simulations were used to simulate the hysteresis loops of the DFL system. Frequency-domain simulations were carried out to calculate the spin-wave dispersion relation of the DFL system. Triangular mesh with a maximum element size of 1 nm inside the ferromagnetic layers and a growth rate of 1.15 outside of the ferromagnetic layers was used.

In the SFL system, the spin-wave dispersion relation was fitted using the analytical formula \cite{gubbiotti1997,moon2013}
\begin{widetext}
\begin{equation}\label{eq:freq}
    f = \frac{\gamma \mu_0}{2\pi} 
    \left(\sqrt{\left(H+M_{\text{S}}\frac{d|k|}{2}+\frac{2 A_\text{{ex}}}{\mu_0 M_{\text{S}}} k^2\right) \left(H+M_{\text{S}}\left(1-\frac{d|k|}{2}\right)+\frac{2 A_{\text{ex}}}{\mu_0 M_{\text{S}}} k^2 - \frac{2 K_u}{\mu_0 M_{\text{S}}}\right)} + \frac{2D}{\mu_0 M_{\text{S}}} k\right).
\end{equation}
\end{widetext}
The frequency difference between the Stokes and anti-Stokes peaks $\Delta f$ can be derived from Eq.~(\ref{eq:freq}) as
\begin{equation}\label{eq:dmi}
    \Delta f = f_{+k} - f_{-k} = \frac{\gamma}{2\pi}\frac{4 D}{M_{\text{S}}} k.
\end{equation}
The field dependence of the frequency was measured at normal-incidence angle $\theta = \SI{0}{\degree}$, therefore, Eq.~(\ref{eq:freq}) simplifies to the following formula
\begin{equation}\label{eq:fmr}
    f = \frac{\gamma \mu_0}{2\pi} \sqrt{H \left(H + M_{\text{S}} - \frac{2 K_{\text{u}}}{\mu_0 M_{\text{S}}} \right)}.
\end{equation}
Basing on Eq.~(\ref{eq:fmr}), the effective magnetization $M_{\text{eff}}$ can be defined as
\begin{equation}\label{eq:ku}
    M_{\text{eff}} = M_{\text{S}} - \frac{2 K_{\text{u}}}{\mu_0 M_{\text{S}}}.
\end{equation}
$M_{\text{S}}$ values are extracted from the out-of-plane magnetization curves obtained by \citet{MENDISCH2019345}. For Co(1.6) is $M_{\text{S}}= \SI{1050}{\kilo\ampere/\meter}$ while for Co(1.95) is $M_{\text{S}}=\SI{1100}{\kilo\ampere/\meter}$.

\section{Results}

\subsection{AGFM hysteresis loops}

\begin{table*}[!t]
    \centering
    \caption{Magnetic parameters of the SFL and DFL samples.}
    \begin{tabular}{c|c|c|c|c|c}
        Abbrev. & Sample & $M_{\text{S}}$ (kA/m) \cite{MENDISCH2019345} & $M_{\text{eff}}$ (kA/m) & $K_{\text{u}}$ (kJ/m$^3$) & $D$ (mJ/m$^2$) \\ \hline
        \textbf{SFL1} & Co(1.95)/W(0.95) & 1100 & 28 & 741 & 0.74 \\ \hline
        \textbf{SFL2} & Co(1.6)/W(0.95) & 1050 & -235 & 848 & 0.84 \\ \hline
        \multirow{2}{*}{\textbf{DFL1}} &\multirow{2}{*}{Co(1.6)/W(0.95)/Co(1.95)} & 1100 & 774 & 225 & -0.64 \\
         & & 1050 & -194 & 821 & 0.72 \\ \hline
        \multirow{2}{*}{\textbf{DFL2}} &\multirow{2}{*}{Co(1.6)/W(0.95)/Co(1.6)} & 1050 & 626 & 280 & -0.49 \\
         & & 1050 & -246 & 855 & 0.84
    \end{tabular}
    \label{tab:params}
\end{table*}

\begin{figure*}[!t]
    \centering
    \includegraphics{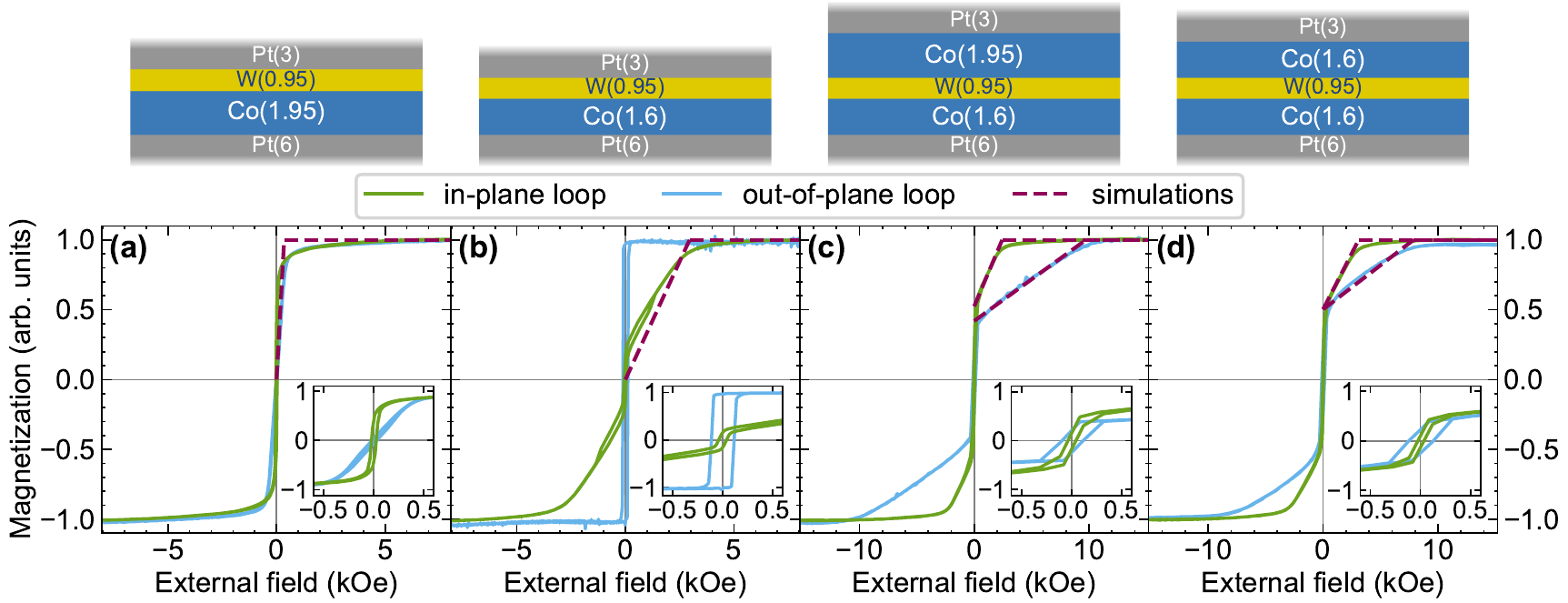}
    \caption{AGFM hysteresis loops measured in the in-plane configuration and out-of-plane configuration. The purple dashed lines represent the fitting using the numerical simulations. In the inset plots, the zoom to low external fields is shown. On top of the figure, the schematic representation of the samples is shown.}
    \label{fig:hysteresis}
\end{figure*}

The AGFM hysteresis loops measured in the film plane (green curves) and perpendicular to the plane (blue curves) are shown in Fig.~\ref{fig:hysteresis}. The results were used to determine $M_{\text{eff}}$ values of the layers in all of the samples. Hard-axis hysteresis loops were fitted with the numerical simulations. The saturation field values obtained from the simulations should be consistent with the $M_{\text{eff}}$ values. The beginning of the magnetization reversal process is sharp, so the extracted $M_{\text{eff}}$ values are considered to be well defined. At this point, the effect of the RKKY interaction between the layers is neglected. Its effect is explained in Section \ref{sec:rkky}. Using the predefined $M_{\text{S}}$ from Ref.~\cite{MENDISCH2019345}, we also determined the PMA constant $K_{\text{u}}$ using Eq.~(\ref{eq:ku}). They are collected in Table~\ref{tab:params}.

In the Co(1.95)/W(0.95) (SFL1) sample [Fig.~\ref{fig:hysteresis}(a)], the parallel loop has a square shape while the perpendicular loop has the typical S-shape behavior characterized by an almost linear dependence on the applied field, indicating an in-plane easy axis. The magnetic saturation in the perpendicular direction is reached in a field much lower than $\mu_0 M_S$, thus indicating that a strong PMA is present in this sample, which competes with the shape anisotropy. $K_u$ is slightly lower than the value required to get a change from an in-plane to an out-of-plane easy magnetization axis, which is 760 kJ/m$^3$, and the sample remains in the in-plane configuration in the remanence. In-plane and perpendicular coercivity values are low, but both loops close at higher field values, i.e., around 0.2--0.3 kOe.

The Co(1.6)/W(0.95) (SFL2) sample [Fig.~\ref{fig:hysteresis}(b)] shows a clear PMA contribution, which dominates over the magnetostatic in-plane contribution, giving rise to an easy magnetization direction perpendicular to the film plane. The hysteresis loop measured with the field applied in the film plane shows a transcritical shape with a saturation field around 3 kOe, low coercivity, but an open loop up to a higher field (approximately 1.2 kOe). The latter feature might depend on a secondary phase with a tilted easy magnetization axis. The hysteresis loop measured in the perpendicular direction shows a square shape with a larger coercive field compared to the in-plane direction. 

\begin{figure*}[!t]
    \centering
    \begin{tabular}{c}
        \includegraphics{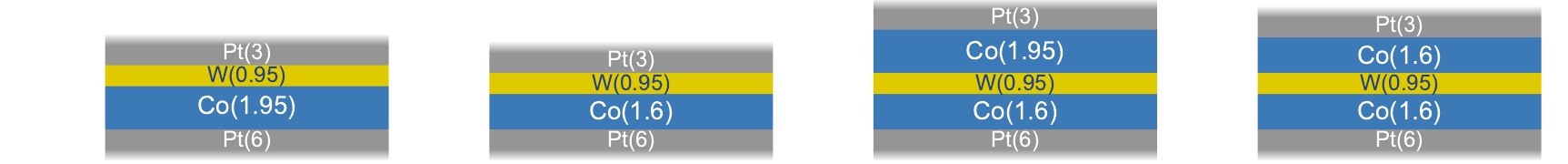} \\
        \includegraphics{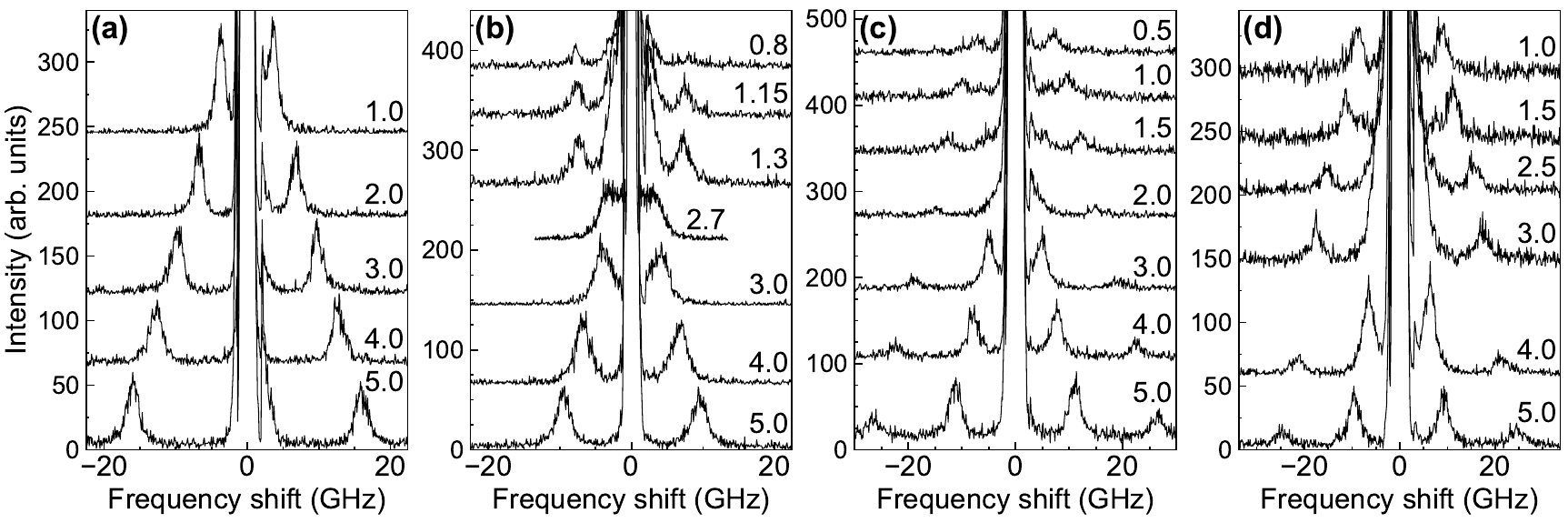} \\
        \includegraphics{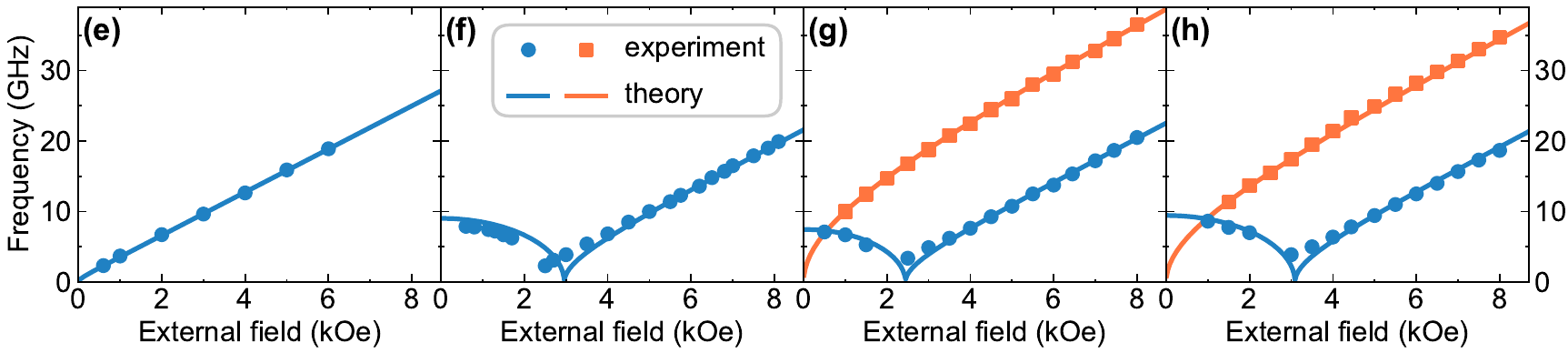}
    \end{tabular}
    \caption{(a-d) The sequence of BLS spectra measured at $k=0$ ($\theta=\SI{0}{\degree}$) and different magnitudes of the externally applied field for all the investigated samples. (e-h) The spin-wave frequency measured by BLS in the function of the external field $H$ and fitting with the theoretical formula in Eq.~(\ref{eq:fmr}).}
    \label{fig:field}
\end{figure*}

In the DFL samples [Figs.~\ref{fig:hysteresis}(c,d)], the PMA contribution is evident. The hysteresis loops along the in-plane and perpendicular directions clearly show the contribution of two magnetic components: one with an easy magnetization axis in the film plane and the other with an easy magnetization axis perpendicular to the film plane. 
In the Co(1.6)/W(0.95)/Co(1.95) (DFL1) sample [Fig.~\ref{fig:hysteresis}(c)], the in-plane easy-axis part appears to have a larger magnetic moment, indicating that it is associated with the Co(1.95) layer, while the out-of-plane easy-axis is present in the Co(1.6) layer. In the Co(1.6)/W(0.95)/Co(1.6) (DFL2) sample [Fig.~\ref{fig:hysteresis}(d)], the in-plane and out-of-plane parts carry the same magnetic moment because of the same thicknesses of the ferromagnetic layers. Also, comparing the magnetic parameters in the DFL and SFL samples, we can ascribe a larger $K_u$ value to the bottom Co layer and smaller $K_u$ to the top Co layer. We associate this effect with the order of the layers deposition. Even though the Co layer has the same vicinity (Pt and W), the effect on the magnetic parameters is different if the material plays the role of the seed or capping layer \cite{chowdhury2012} as well as depends on the growth conditions \cite{Mallick_2018}. Moreover, in both DFL samples, the coercive field values are larger in the perpendicular direction.

\subsection{Field-dependent BLS measurements}

Figs.~\ref{fig:field}(a-d) present the sequences of measured BLS spectra and Figs.~\ref{fig:field}(e-h) present the frequency values plotted as a function of $H$ at $k=0$ for all of the investigated samples. BLS spectra were measured starting from $H = 8$ kOe and decreasing it down to zero, i.e., along the descending branches of the in-plane loops in Fig.~\ref{fig:hysteresis}. For SFL samples, only one peak is observed. For the SFL1 sample, it has a monotonic dependence on $H$, while for the SFL2 sample, it first decreases, reaching a minimum at 2.4 kOe, and then increases again. For DFL samples, two peaks are visible on both the Stokes and anti-Stokes sides of the spectra. Here, the field dependence is more complicated since the frequency of high-frequency mode monotonously decreases with decreasing $H$ while the low-frequency peak first decreases, reaches a minimum, and then increases again. Since the spectra are measured at $k=0$, the frequency position of the peaks are symmetric ($f_{\text{S}} = f_{\text{AS}}$) on both the Stokes and anti-Stokes sides of the spectra.. Basing on PMA constants derived from the AGFM hysteresis loops, we used Eq.~(\ref{eq:fmr}) to determine the gyromagnetic ratio $\gamma$ of the analyzed samples. We reached satisfying fitting for $\gamma = \SI{192}{\radian/(\tesla\cdot\second)}$, as shown in Figs.~\ref{fig:field}(e-h) with solid lines. These results, together with the values of the saturation fields derived from the AGFM measured loops, permit to unambiguously affirm that above 4 kOe, the magnetization is saturated and aligned parallel to the sample plane. Therefore, we performed the wavevector-dependent measurements for an externally applied field of $\pm 5.5$ kOe.

\subsection{Role of RKKY interaction in DFL samples}\label{sec:rkky}

The numerical simulations were used to check the presence of the RKKY interaction in the DFL samples. First, $K_{\text{u},1}$ and $K_{\text{u},2}$ were fitted to the hysteresis loops from Fig.~\ref{fig:hysteresis} in the function of the RKKY constant $J$. The results show that the difference between $K_{\text{u},1}$ and $K_{\text{u},2}$ increases with the RKKY constant. However, the hysteresis loop does not change its character. The shape is identical to the shape of the loop of the noninteracting layers. The possible cause is the large difference between $M_{\text{eff}}$. In the next step, the RKKY constant (along with related to it $K_{\text{u},1}$ and $K_{\text{u},2}$) was fitted to the field dependence of the frequency results obtained from the BLS measurements shown in Figs.~\ref{fig:field}(e-h). The gyromagnetic ratio $\gamma$ was used to fit the low-frequency mode, and the RKKY constant was fitted to the high-frequency mode. The value of the RKKY constant for DFL1 equals $\SI{-100}{\micro \joule / \meter^2}$ for $\gamma = \SI{195}{\radian/(\tesla\cdot\second)}$ and for DFL2 equals $\SI{20}{\micro \joule / \meter^2}$ for $\gamma = \SI{190}{\radian/(\tesla\cdot\second)}$. However, the best fit assuming the absence of the RKKY is as far as 1 GHz from the satisfactory fit. It lies within the error range, which considers the reading of $M_{\rm{eff}}$ from the hysteresis loops and instrument errors. In general, the frequency shift is small even for a large value of the RKKY constant (exceeding $\pm\SI{100}{\micro \joule / \meter^2}$). Moreover, the shape of the hysteresis loops and the field dependence of the frequency with RKKY are identical to the case of the noninteracting layers. Therefore, the presence of the RKKY interaction can not be confirmed when the difference between $M_{\text{eff}}$ of the layers in the bilayered structure is too large, and thus the RKKY interaction in DFL samples is assumed to be negligible.

\subsection{Wavevector-dependent BLS measurements}

\begin{figure*}[!t]
    \centering
    \begin{tabular}{c}
        \includegraphics{schemas3_multi.pdf} \\
        \includegraphics{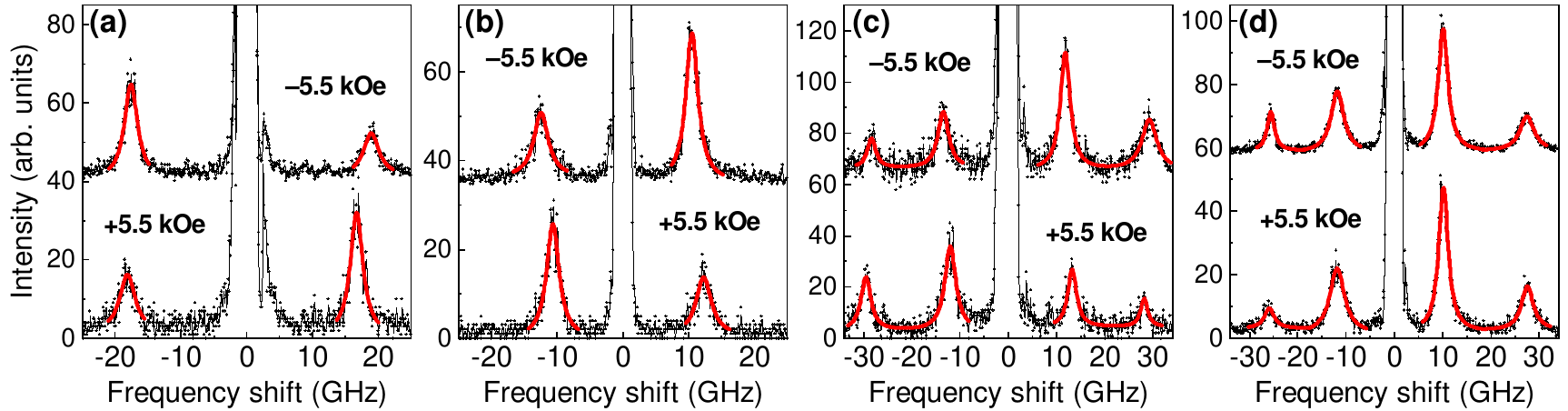} \\
        \includegraphics{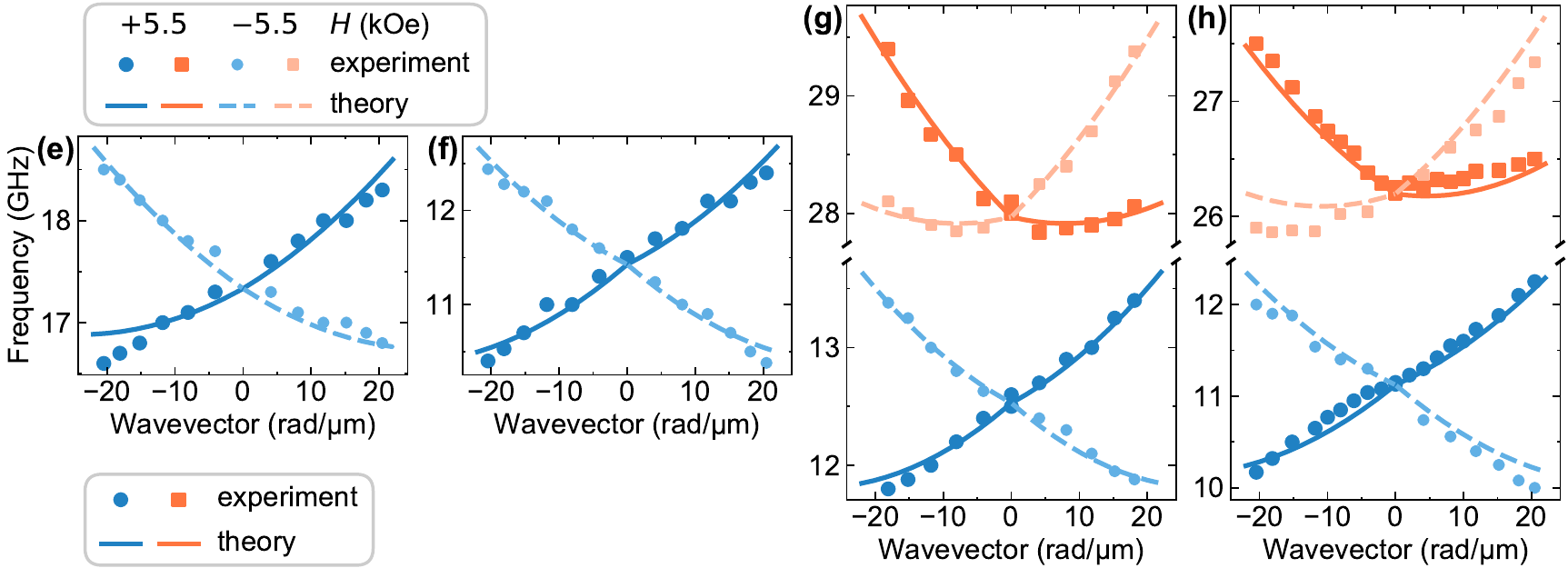} \\
        \includegraphics{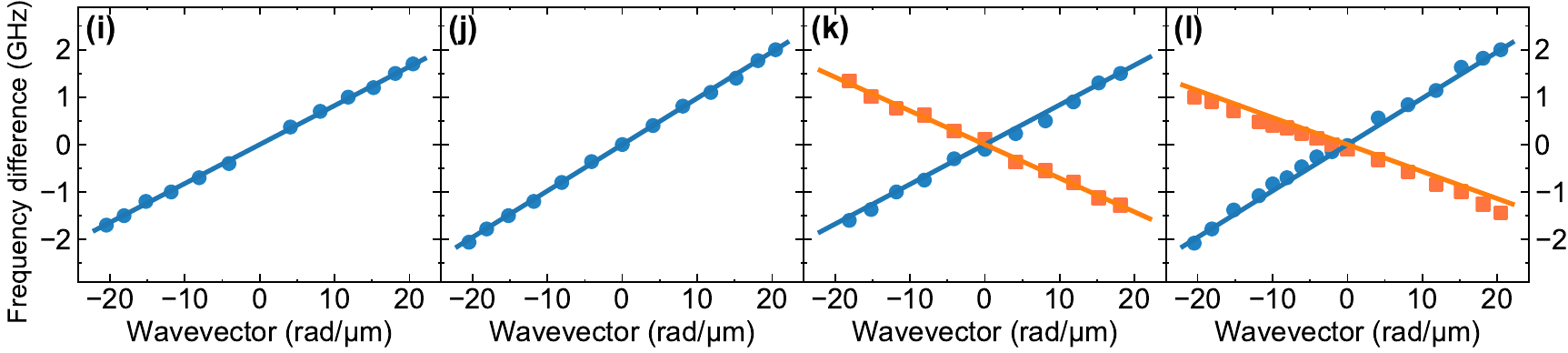}
    \end{tabular}
    \caption{(a-d) Measured BLS spectra at $k = \SI{18.1}{\radian/\micro\meter}$ ($\theta = \SI{50}{\degree}$) and for two different orientation of the external field: $H= +5.5$ and $-5.5$ kOe. Black points represent the measured spectra, while red curves are the Lorentzian fitting of the peaks. (e-h) Dispersion relations measured by the BLS spectroscopy in the external field of $\pm5.5$ kOe and fitted using finite-element method simulations. (i-l) Frequency asymmetry plots measured by the BLS spectroscopy with the linear regression fitting basing on Eq.~(\ref{eq:dmi}).}
    \label{fig:disp}
\end{figure*}

In order to derive the iDMI constant, we measured the BLS spectra for different wavevector values in the range from 0 to $\SI{20}{\radian/\micro\meter}$ for $H$ fixed at $+5.5$ and $-5.5$ kOe. Typical BLS spectra measured at these fields for $k = \SI{18.1}{\radian/\micro\meter}$ are shown in Fig.~\ref{fig:disp}(a-d) for all the investigated samples. Because of the small sample thickness, both the Stokes and anti-Stokes peaks, corresponding to spin waves propagating in opposite directions, are simultaneously observed with slightly different intensity. Frequencies of the spin-wave modes were extracted by fitting the peaks in the BLS spectra (displayed in Figs.~\ref{fig:disp}(a-d) by the red curves) with Lorentzian functions and were plotted as a function of their respective wavevectors $k$ in Figs.~\ref{fig:disp}(e-h). It is noteworthy that on reversing the direction of the applied magnetic field, the respective center frequencies and intensities of the Stokes and anti-Stokes peaks interchanges due to the fact that reversing the magnetization is equivalent to a reversal operation. For that reason, the following discussion will pertain to the results in positive external magnetic field. The dispersions were calculated by the finite-element method simulations in COMSOL Multiphysics. We set the exchange stiffness $A_{\text{ex}} = \SI{15}{\pico\joule/\meter}$. The values of $M_{\text{S}}$ and $K_u$ used in simulations are written in Table~\ref{tab:params}. The iDMI constant $D$ was the free parameter in the fitting procedure for each sample separately.

In the SFL samples [Figs.~\ref{fig:disp}(e,f)], the frequency dependence is close to linear and has a positive slope. A more detailed analysis of Eq.~(\ref{eq:freq}) allows explaining the character of the dispersion relation. First, the effect of the dipolar interaction is significantly weakened by the PMA as $M_{\text{S}} \approx 2 K_u/\mu_0 M_{\text{S}}$ and so $M_{\text{eff}} \approx 0$. In fact, when $M_{\text{eff}} < 0$, the surface wave propagates backward, and the dispersion relation has a negative slope even with the absence of iDMI, until the effect of the exchange interaction starts to dominate. The small thickness of the layer leads to the additional flattening of the dispersion. In the BLS range of the wavevector, the effect of the exchange interaction is small, and so the dispersion relation is close to linear. The slope of the dispersion is directly connected to the iDMI.

The dispersion relation of the DFL samples [Figs.~\ref{fig:disp}(g,h)] shows two branches of a different character. Comparing the dispersion relations with the systems of noninteracting layers, the frequency is changing by less than 100 MHz, indicating very weak dipolar coupling between the layers. The weakening of the dipolar coupling is caused by the large difference between the PMA of the layers. Interestingly, it allows us to use the approximation of the noninteracting layers, even for separations as small as 1 nm. The low-frequency mode is close to linear with the positive slope. Its character is identical to the single-layer dispersion confirming the connection with the bottom layer. The high-frequency mode is flat on the positive wavevectors' side and has a negative slope on the negative wavevectors' side. The flattening is the result of the compensation of dynamic dipolar interaction by the iDMI.

Figs.~\ref{fig:disp}(i-l) shows the frequency difference $\Delta f$ between the Stokes and the anti-Stokes peaks measured as a function of the wavevector $k$. The values for positive wavevectors represents the measurements in the positive external field and negative wavevectors -- in the negative external field. The frequency asymmetry exhibits a linear dependence as a function of $k$, agreeing with the theoretical prediction of Eq.~(\ref{eq:dmi}). The frequency difference plots were fitted using the linear regression method based on Eq.~(\ref{eq:dmi}), and the values of iDMI constant $D$ are collected in Table~\ref{tab:params}. The values are in agreement with values presented in the literature for the similar thickness of Co layer \cite{cho2015thickness}, where the iDMI constant is about $\SI{1}{\milli \joule / \meter^2}$ for $ \SI{1.6}{\nano\meter}$ Co thickness and lower for larger thicknesses of Co.

In the DFL samples [Figs.~\ref{fig:disp}(k,l)], the frequency asymmetry always has a linear dependence on $k$, but for the two modes, it has an opposite sign since the Co layers have Pt as a bottom or top layer. However, the iDMI constants in the layers differ. This effect can come from the different thicknesses of the Pt layers, as well as, in the DFL1 sample [Fig.~\ref{fig:disp}(k)], from the different thicknesses of the Co layers. The sputtering order of the layers can also affect the iDMI constant absolute value, as it is observed for the PMA constant.

\section{Conclusions}

We have studied the iDMI and spin-wave dynamics in Pt/Co/W/Co/Pt DFL structures with the opposite arrangement of Co and Pt layers and Pt/Co/W SFL samples. AGFM hysteresis loops show a strong PMA contribution, leading to a resultant out-of-plane easy magnetization axis in Co(1.6) layer and in-plane easy magnetization axis in Co(1.95) layer with small effective magnetization. Hysteresis loops of DFL samples indicate both in-plane and out-of-plane anisotropy contributions, also pointing at a significant difference between the PMA constants of the Co layers. Field-dependent measurements of the BLS spectra at normal incidence confirm the presence of strong PMA in the Co layers. Numerical simulations were made to check the presence of the RKKY interaction through the W spacer. We find that AGFM and BLS results can be fitted without this interaction taken into account, which does not confirm the presence of the RKKY interaction in the DFL samples. 

Wavevector-dependent BLS measurements were made in a large external field of $\pm 5.5$ kOe to reach the in-plane configuration of magnetization. The results show the iDMI-dominated linear dispersion in SFL samples and two-mode dispersion in DFL samples related to the two Co layers in the studied system. The strong asymmetry between the Stokes and anti-Stokes peaks is present. In the DFL samples, the slope of the two branches and thus the signs of the iDMI constant are opposite due to the opposite arrangement of Co and Pt layers. The iDMI constant values vary between 0.49 and \SI{0.84}{\milli\joule/\meter^2}, being in line with the values known from the literature. 

In DFL samples, the linear low-frequency mode is dominated by iDMI and resembles the dispersions measured in SFLs. The asymmetric high-frequency mode consists of a flat branch for positive wavevectors and a strongly dispersive branch for the negative wavevectors at the selected orientation of the external magnetic field. A comparison with the system of noninteracting layers points out a significant weakening of the interlayer coupling via the dynamic dipolar field. This uncoupling of spin waves between the Co layers separated by only 0.95-nm-thick nonmagnetic spacer arises from the significantly different frequencies of oscillations in both layers, with the difference equaling 15.5 GHz at $k=0$ for both DFL1 and DFL2 samples. Such a significant difference in frequencies in nominally very similar Co films is attributed to the different PMAs originating in different deposition sequences for the bottom and top Co layers.

To sum up, we design the system based on ultrathin layers composed of the same set of materials, operating at two frequency ranges, both possessing nonreciprocal spin-wave dispersion relations but with different characteristics, indicating interesting properties for magnonics, especially for the design of the multifunctional spin-wave devices. 

\section*{Acknowledgements}

The authors thank Maciej Krawczyk for a fruitful discussion. K.S. acknowledges the financial support from the National Science Center of Poland, project no. \mbox{UMO-2018/30/Q/ST3/00416}. S.M. and M.B. acknowledge the financial support by the German Research Foundation (DFG), priority program SPP2137 project no. 403505866, as well as the TUM International Graduate School of Science and Engineering (IGSSE), project no. GSC 81-24184165. M.M. acknowledges funding from the Slovak Grant Agency APVV, no. \mbox{APVV-16-0068} (NanoSky) and \mbox{APVV-19-0311} (RSWFA). This study was performed in the frame of the implementation of the project Building-up Centre for Advanced Materials Application of the Slovak Academy of Sciences, ITMS project code 313021T081 supported by Research \& Innovation Operational Programme funded by the ERDF. G.G. acknowledges the financial support by the European Metrology Programme for Innovation and Research (EMPIR) under the Grant Agreement 17FUN08 TOPS.

\bibliography{main} 

\begin{thebibliography}{48}%
\makeatletter
\providecommand \@ifxundefined [1]{%
 \@ifx{#1\undefined}
}%
\providecommand \@ifnum [1]{%
 \ifnum #1\expandafter \@firstoftwo
 \else \expandafter \@secondoftwo
 \fi
}%
\providecommand \@ifx [1]{%
 \ifx #1\expandafter \@firstoftwo
 \else \expandafter \@secondoftwo
 \fi
}%
\providecommand \natexlab [1]{#1}%
\providecommand \enquote  [1]{``#1''}%
\providecommand \bibnamefont  [1]{#1}%
\providecommand \bibfnamefont [1]{#1}%
\providecommand \citenamefont [1]{#1}%
\providecommand \href@noop [0]{\@secondoftwo}%
\providecommand \href [0]{\begingroup \@sanitize@url \@href}%
\providecommand \@href[1]{\@@startlink{#1}\@@href}%
\providecommand \@@href[1]{\endgroup#1\@@endlink}%
\providecommand \@sanitize@url [0]{\catcode `\\12\catcode `\$12\catcode
  `\&12\catcode `\#12\catcode `\^12\catcode `\_12\catcode `\%12\relax}%
\providecommand \@@startlink[1]{}%
\providecommand \@@endlink[0]{}%
\providecommand \url  [0]{\begingroup\@sanitize@url \@url }%
\providecommand \@url [1]{\endgroup\@href {#1}{\urlprefix }}%
\providecommand \urlprefix  [0]{URL }%
\providecommand \Eprint [0]{\href }%
\providecommand \doibase [0]{https://doi.org/}%
\providecommand \selectlanguage [0]{\@gobble}%
\providecommand \bibinfo  [0]{\@secondoftwo}%
\providecommand \bibfield  [0]{\@secondoftwo}%
\providecommand \translation [1]{[#1]}%
\providecommand \BibitemOpen [0]{}%
\providecommand \bibitemStop [0]{}%
\providecommand \bibitemNoStop [0]{.\EOS\space}%
\providecommand \EOS [0]{\spacefactor3000\relax}%
\providecommand \BibitemShut  [1]{\csname bibitem#1\endcsname}%
\let\auto@bib@innerbib\@empty
\bibitem [{\citenamefont {Moriya}(1960)}]{Moriya60}%
  \BibitemOpen
  \bibfield  {author} {\bibinfo {author} {\bibfnamefont {T.}~\bibnamefont
  {Moriya}},\ }\bibfield  {title} {\bibinfo {title} {Anisotropic superexchange
  interaction and weak ferromagnetism},\ }\href@noop {} {\bibfield  {journal}
  {\bibinfo  {journal} {Phys. Rev.}\ }\textbf {\bibinfo {volume} {120}},\
  \bibinfo {pages} {91} (\bibinfo {year} {1960})}\BibitemShut {NoStop}%
\bibitem [{\citenamefont {Finocchio}\ \emph {et~al.}(2016)\citenamefont
  {Finocchio}, \citenamefont {Büttner}, \citenamefont {Tomasello},
  \citenamefont {Carpentieri},\ and\ \citenamefont {Kläui}}]{Finocchio16}%
  \BibitemOpen
  \bibfield  {author} {\bibinfo {author} {\bibfnamefont {G.}~\bibnamefont
  {Finocchio}}, \bibinfo {author} {\bibfnamefont {F.}~\bibnamefont {Büttner}},
  \bibinfo {author} {\bibfnamefont {R.}~\bibnamefont {Tomasello}}, \bibinfo
  {author} {\bibfnamefont {M.}~\bibnamefont {Carpentieri}},\ and\ \bibinfo
  {author} {\bibfnamefont {M.}~\bibnamefont {Kläui}},\ }\bibfield  {title}
  {\bibinfo {title} {Magnetic skyrmions: from fundamental to applications},\
  }\href@noop {} {\bibfield  {journal} {\bibinfo  {journal} {J. Phys. D: Appl.
  Phys.}\ }\textbf {\bibinfo {volume} {49}},\ \bibinfo {pages} {423001}
  (\bibinfo {year} {2016})}\BibitemShut {NoStop}%
\bibitem [{\citenamefont {dos Santos}\ \emph
  {et~al.}(2020{\natexlab{a}})\citenamefont {dos Santos}, \citenamefont {dos
  Santos~Dias},\ and\ \citenamefont {Lounis}}]{santos2020nonreciprocity}%
  \BibitemOpen
  \bibfield  {author} {\bibinfo {author} {\bibfnamefont {F.~J.}\ \bibnamefont
  {dos Santos}}, \bibinfo {author} {\bibfnamefont {M.}~\bibnamefont {dos
  Santos~Dias}},\ and\ \bibinfo {author} {\bibfnamefont {S.}~\bibnamefont
  {Lounis}},\ }\bibfield  {title} {\bibinfo {title} {Nonreciprocity of spin
  waves in noncollinear magnets due to the Dzyaloshinskii-Moriya interaction},\
  }\href {https://doi.org/10.1103/PhysRevB.102.104401} {\bibfield  {journal}
  {\bibinfo  {journal} {Phys. Rev. B}\ }\textbf {\bibinfo {volume} {102}},\
  \bibinfo {pages} {104401} (\bibinfo {year} {2020}{\natexlab{a}})}\BibitemShut
  {NoStop}%
\bibitem [{\citenamefont {Udvardi}\ and\ \citenamefont
  {Szunyogh}(2009)}]{udvardi2009}%
  \BibitemOpen
  \bibfield  {author} {\bibinfo {author} {\bibfnamefont {L.}~\bibnamefont
  {Udvardi}}\ and\ \bibinfo {author} {\bibfnamefont {L.}~\bibnamefont
  {Szunyogh}},\ }\bibfield  {title} {\bibinfo {title} {Chiral asymmetry of the
  spin-wave spectra in ultrathin magnetic films},\ }\href@noop {} {\bibfield
  {journal} {\bibinfo  {journal} {Phys. Rev. Lett.}\ }\textbf {\bibinfo
  {volume} {102}},\ \bibinfo {pages} {207204} (\bibinfo {year}
  {2009})}\BibitemShut {NoStop}%
\bibitem [{\citenamefont {Cort{\'{e}}s-Ortu{\~{n}}o}\ and\ \citenamefont
  {Landeros}(2013)}]{cortes-ortuno2013}%
  \BibitemOpen
  \bibfield  {author} {\bibinfo {author} {\bibfnamefont {D.}~\bibnamefont
  {Cort{\'{e}}s-Ortu{\~{n}}o}}\ and\ \bibinfo {author} {\bibfnamefont
  {P.}~\bibnamefont {Landeros}},\ }\bibfield  {title} {\bibinfo {title}
  {Influence of the Dzyaloshinskii{\textendash}Moriya interaction on the
  spin-wave spectra of thin films},\ }\href@noop {} {\bibfield  {journal}
  {\bibinfo  {journal} {J. Phys. Condens. Matter}\ }\textbf {\bibinfo {volume}
  {25}},\ \bibinfo {pages} {156001} (\bibinfo {year} {2013})}\BibitemShut
  {NoStop}%
\bibitem [{\citenamefont {Mruczkiewicz}\ and\ \citenamefont
  {Krawczyk}(2016)}]{mruczkiewicz2016}%
  \BibitemOpen
  \bibfield  {author} {\bibinfo {author} {\bibfnamefont {M.}~\bibnamefont
  {Mruczkiewicz}}\ and\ \bibinfo {author} {\bibfnamefont {M.}~\bibnamefont
  {Krawczyk}},\ }\bibfield  {title} {\bibinfo {title} {Influence of the
  Dzyaloshinskii-Moriya interaction on the FMR spectrum of magnonic crystals
  and confined structures},\ }\href@noop {} {\bibfield  {journal} {\bibinfo
  {journal} {Phys. Rev. B}\ }\textbf {\bibinfo {volume} {94}},\ \bibinfo
  {pages} {024434} (\bibinfo {year} {2016})}\BibitemShut {NoStop}%
\bibitem [{\citenamefont {Szulc}\ \emph {et~al.}(2020)\citenamefont {Szulc},
  \citenamefont {Graczyk}, \citenamefont {Mruczkiewicz}, \citenamefont
  {Gubbiotti},\ and\ \citenamefont {Krawczyk}}]{szulc2020spin}%
  \BibitemOpen
  \bibfield  {author} {\bibinfo {author} {\bibfnamefont {K.}~\bibnamefont
  {Szulc}}, \bibinfo {author} {\bibfnamefont {P.}~\bibnamefont {Graczyk}},
  \bibinfo {author} {\bibfnamefont {M.}~\bibnamefont {Mruczkiewicz}}, \bibinfo
  {author} {\bibfnamefont {G.}~\bibnamefont {Gubbiotti}},\ and\ \bibinfo
  {author} {\bibfnamefont {M.}~\bibnamefont {Krawczyk}},\ }\bibfield  {title}
  {\bibinfo {title} {Spin-wave diode and circulator based on unidirectional
  coupling},\ }\href@noop {} {\bibfield  {journal} {\bibinfo  {journal} {Phys.
  Rev. Appl.}\ }\textbf {\bibinfo {volume} {14}},\ \bibinfo {pages} {034063}
  (\bibinfo {year} {2020})}\BibitemShut {NoStop}%
\bibitem [{\citenamefont {Belmeguenai}\ \emph {et~al.}(2015)\citenamefont
  {Belmeguenai}, \citenamefont {Adam}, \citenamefont {Roussign\'e},
  \citenamefont {Eimer}, \citenamefont {Devolder}, \citenamefont {Kim},
  \citenamefont {Cherif}, \citenamefont {Stashkevich},\ and\ \citenamefont
  {Thiaville}}]{belmeguenai2015}%
  \BibitemOpen
  \bibfield  {author} {\bibinfo {author} {\bibfnamefont {M.}~\bibnamefont
  {Belmeguenai}}, \bibinfo {author} {\bibfnamefont {J.-P.}\ \bibnamefont
  {Adam}}, \bibinfo {author} {\bibfnamefont {Y.}~\bibnamefont {Roussign\'e}},
  \bibinfo {author} {\bibfnamefont {S.}~\bibnamefont {Eimer}}, \bibinfo
  {author} {\bibfnamefont {T.}~\bibnamefont {Devolder}}, \bibinfo {author}
  {\bibfnamefont {J.-V.}\ \bibnamefont {Kim}}, \bibinfo {author} {\bibfnamefont
  {S.~M.}\ \bibnamefont {Cherif}}, \bibinfo {author} {\bibfnamefont
  {A.}~\bibnamefont {Stashkevich}},\ and\ \bibinfo {author} {\bibfnamefont
  {A.}~\bibnamefont {Thiaville}},\ }\bibfield  {title} {\bibinfo {title}
  {Interfacial Dzyaloshinskii-Moriya interaction in perpendicularly magnetized
  ${\text{Pt/Co/AlO}}_{x}$ ultrathin films measured by Brillouin light
  spectroscopy},\ }\href@noop {} {\bibfield  {journal} {\bibinfo  {journal}
  {Phys. Rev. B}\ }\textbf {\bibinfo {volume} {91}},\ \bibinfo {pages} {180405}
  (\bibinfo {year} {2015})}\BibitemShut {NoStop}%
\bibitem [{\citenamefont {Cho}\ \emph {et~al.}(2015)\citenamefont {Cho},
  \citenamefont {Kim}, \citenamefont {Lee}, \citenamefont {Kim}, \citenamefont
  {Lavrijsen}, \citenamefont {Solignac}, \citenamefont {Yin}, \citenamefont
  {Han}, \citenamefont {Van~Hoof}, \citenamefont {Swagten}, \citenamefont
  {Koopmans},\ and\ \citenamefont {You}}]{cho2015thickness}%
  \BibitemOpen
  \bibfield  {author} {\bibinfo {author} {\bibfnamefont {J.}~\bibnamefont
  {Cho}}, \bibinfo {author} {\bibfnamefont {N.-H.}\ \bibnamefont {Kim}},
  \bibinfo {author} {\bibfnamefont {S.}~\bibnamefont {Lee}}, \bibinfo {author}
  {\bibfnamefont {J.-S.}\ \bibnamefont {Kim}}, \bibinfo {author} {\bibfnamefont
  {R.}~\bibnamefont {Lavrijsen}}, \bibinfo {author} {\bibfnamefont
  {A.}~\bibnamefont {Solignac}}, \bibinfo {author} {\bibfnamefont
  {Y.}~\bibnamefont {Yin}}, \bibinfo {author} {\bibfnamefont {D.-S.}\
  \bibnamefont {Han}}, \bibinfo {author} {\bibfnamefont {N.~J.}\ \bibnamefont
  {Van~Hoof}}, \bibinfo {author} {\bibfnamefont {H.~J.}\ \bibnamefont
  {Swagten}}, \bibinfo {author} {\bibfnamefont {B.}~\bibnamefont {Koopmans}},\
  and\ \bibinfo {author} {\bibfnamefont {C.-Y.}\ \bibnamefont {You}},\
  }\bibfield  {title} {\bibinfo {title} {Thickness dependence of the
  interfacial Dzyaloshinskii--Moriya interaction in inversion symmetry broken
  systems},\ }\href@noop {} {\bibfield  {journal} {\bibinfo  {journal} {Nat.
  Commun.}\ }\textbf {\bibinfo {volume} {6}},\ \bibinfo {pages} {1} (\bibinfo
  {year} {2015})}\BibitemShut {NoStop}%
\bibitem [{\citenamefont {Boulle}\ \emph {et~al.}(2016)\citenamefont {Boulle},
  \citenamefont {Vogel}, \citenamefont {Yang}, \citenamefont {Pizzini},
  \citenamefont {de~Souza~Chaves}, \citenamefont {Locatelli}, \citenamefont
  {Mente{\c{s}}}, \citenamefont {Sala}, \citenamefont {Buda-Prejbeanu},
  \citenamefont {Klein}, \citenamefont {Belmeguenai}, \citenamefont
  {Roussigné}, \citenamefont {Stashkevich}, \citenamefont {Chérif},
  \citenamefont {Aballe}, \citenamefont {Foerster}, \citenamefont {Chshiev},
  \citenamefont {Auffret}, \citenamefont {Miron},\ and\ \citenamefont
  {Gaudin}}]{boulle2016room}%
  \BibitemOpen
  \bibfield  {author} {\bibinfo {author} {\bibfnamefont {O.}~\bibnamefont
  {Boulle}}, \bibinfo {author} {\bibfnamefont {J.}~\bibnamefont {Vogel}},
  \bibinfo {author} {\bibfnamefont {H.}~\bibnamefont {Yang}}, \bibinfo {author}
  {\bibfnamefont {S.}~\bibnamefont {Pizzini}}, \bibinfo {author} {\bibfnamefont
  {D.}~\bibnamefont {de~Souza~Chaves}}, \bibinfo {author} {\bibfnamefont
  {A.}~\bibnamefont {Locatelli}}, \bibinfo {author} {\bibfnamefont {T.~O.}\
  \bibnamefont {Mente{\c{s}}}}, \bibinfo {author} {\bibfnamefont
  {A.}~\bibnamefont {Sala}}, \bibinfo {author} {\bibfnamefont {L.~D.}\
  \bibnamefont {Buda-Prejbeanu}}, \bibinfo {author} {\bibfnamefont
  {O.}~\bibnamefont {Klein}}, \bibinfo {author} {\bibfnamefont
  {M.}~\bibnamefont {Belmeguenai}}, \bibinfo {author} {\bibfnamefont
  {Y.}~\bibnamefont {Roussigné}}, \bibinfo {author} {\bibfnamefont
  {A.}~\bibnamefont {Stashkevich}}, \bibinfo {author} {\bibfnamefont {S.~M.}\
  \bibnamefont {Chérif}}, \bibinfo {author} {\bibfnamefont {L.}~\bibnamefont
  {Aballe}}, \bibinfo {author} {\bibfnamefont {M.}~\bibnamefont {Foerster}},
  \bibinfo {author} {\bibfnamefont {M.}~\bibnamefont {Chshiev}}, \bibinfo
  {author} {\bibfnamefont {S.}~\bibnamefont {Auffret}}, \bibinfo {author}
  {\bibfnamefont {I.~M.}\ \bibnamefont {Miron}},\ and\ \bibinfo {author}
  {\bibfnamefont {G.}~\bibnamefont {Gaudin}},\ }\bibfield  {title} {\bibinfo
  {title} {Room-temperature chiral magnetic skyrmions in ultrathin magnetic
  nanostructures},\ }\href@noop {} {\bibfield  {journal} {\bibinfo  {journal}
  {Nat. Nanotechnol.}\ }\textbf {\bibinfo {volume} {11}},\ \bibinfo {pages}
  {449} (\bibinfo {year} {2016})}\BibitemShut {NoStop}%
\bibitem [{\citenamefont {Woo}\ \emph {et~al.}(2016)\citenamefont {Woo},
  \citenamefont {Litzius}, \citenamefont {Kr{\"u}ger}, \citenamefont {Im},
  \citenamefont {Caretta}, \citenamefont {Richter}, \citenamefont {Mann},
  \citenamefont {Krone}, \citenamefont {Reeve}, \citenamefont {Weigand},
  \citenamefont {Agrawal}, \citenamefont {Lemesh}, \citenamefont {Mawass},
  \citenamefont {Fischer}, \citenamefont {Kläui},\ and\ \citenamefont
  {Beach}}]{woo2016observation}%
  \BibitemOpen
  \bibfield  {author} {\bibinfo {author} {\bibfnamefont {S.}~\bibnamefont
  {Woo}}, \bibinfo {author} {\bibfnamefont {K.}~\bibnamefont {Litzius}},
  \bibinfo {author} {\bibfnamefont {B.}~\bibnamefont {Kr{\"u}ger}}, \bibinfo
  {author} {\bibfnamefont {M.-Y.}\ \bibnamefont {Im}}, \bibinfo {author}
  {\bibfnamefont {L.}~\bibnamefont {Caretta}}, \bibinfo {author} {\bibfnamefont
  {K.}~\bibnamefont {Richter}}, \bibinfo {author} {\bibfnamefont
  {M.}~\bibnamefont {Mann}}, \bibinfo {author} {\bibfnamefont {A.}~\bibnamefont
  {Krone}}, \bibinfo {author} {\bibfnamefont {R.~M.}\ \bibnamefont {Reeve}},
  \bibinfo {author} {\bibfnamefont {M.}~\bibnamefont {Weigand}}, \bibinfo
  {author} {\bibfnamefont {P.}~\bibnamefont {Agrawal}}, \bibinfo {author}
  {\bibfnamefont {I.}~\bibnamefont {Lemesh}}, \bibinfo {author} {\bibfnamefont
  {M.-A.}\ \bibnamefont {Mawass}}, \bibinfo {author} {\bibfnamefont
  {P.}~\bibnamefont {Fischer}}, \bibinfo {author} {\bibfnamefont
  {M.}~\bibnamefont {Kläui}},\ and\ \bibinfo {author} {\bibfnamefont
  {G.~S.~D.}\ \bibnamefont {Beach}},\ }\bibfield  {title} {\bibinfo {title}
  {Observation of room-temperature magnetic skyrmions and their current-driven
  dynamics in ultrathin metallic ferromagnets},\ }\href@noop {} {\bibfield
  {journal} {\bibinfo  {journal} {Nat. Mater.}\ }\textbf {\bibinfo {volume}
  {15}},\ \bibinfo {pages} {501} (\bibinfo {year} {2016})}\BibitemShut
  {NoStop}%
\bibitem [{\citenamefont {Zhang}\ \emph {et~al.}(2017)\citenamefont {Zhang},
  \citenamefont {Cao}, \citenamefont {Qiao}, \citenamefont {Tang},
  \citenamefont {Cao}, \citenamefont {Zhao}, \citenamefont {Eimer},
  \citenamefont {Si}, \citenamefont {Lei}, \citenamefont {Wang}, \citenamefont
  {Lin}, \citenamefont {Zhang}, \citenamefont {Wu},\ and\ \citenamefont
  {Zhao}}]{zhang2017}%
  \BibitemOpen
  \bibfield  {author} {\bibinfo {author} {\bibfnamefont {B.}~\bibnamefont
  {Zhang}}, \bibinfo {author} {\bibfnamefont {A.}~\bibnamefont {Cao}}, \bibinfo
  {author} {\bibfnamefont {J.}~\bibnamefont {Qiao}}, \bibinfo {author}
  {\bibfnamefont {M.}~\bibnamefont {Tang}}, \bibinfo {author} {\bibfnamefont
  {K.}~\bibnamefont {Cao}}, \bibinfo {author} {\bibfnamefont {X.}~\bibnamefont
  {Zhao}}, \bibinfo {author} {\bibfnamefont {S.}~\bibnamefont {Eimer}},
  \bibinfo {author} {\bibfnamefont {Z.}~\bibnamefont {Si}}, \bibinfo {author}
  {\bibfnamefont {N.}~\bibnamefont {Lei}}, \bibinfo {author} {\bibfnamefont
  {Z.}~\bibnamefont {Wang}}, \bibinfo {author} {\bibfnamefont {X.}~\bibnamefont
  {Lin}}, \bibinfo {author} {\bibfnamefont {Z.}~\bibnamefont {Zhang}}, \bibinfo
  {author} {\bibfnamefont {M.}~\bibnamefont {Wu}},\ and\ \bibinfo {author}
  {\bibfnamefont {W.}~\bibnamefont {Zhao}},\ }\bibfield  {title} {\bibinfo
  {title} {Influence of heavy metal materials on magnetic properties of
  Pt/Co/heavy metal tri-layered structures},\ }\href
  {https://doi.org/10.1063/1.4973477} {\bibfield  {journal} {\bibinfo
  {journal} {Appl. Phys. Lett.}\ }\textbf {\bibinfo {volume} {110}},\ \bibinfo
  {pages} {012405} (\bibinfo {year} {2017})}\BibitemShut {NoStop}%
\bibitem [{\citenamefont {Yu}\ \emph {et~al.}(2016)\citenamefont {Yu},
  \citenamefont {Qiu}, \citenamefont {Wu}, \citenamefont {Yoon}, \citenamefont
  {Deorani}, \citenamefont {Besbas}, \citenamefont {Manchon},\ and\
  \citenamefont {Yang}}]{yu2016spin}%
  \BibitemOpen
  \bibfield  {author} {\bibinfo {author} {\bibfnamefont {J.}~\bibnamefont
  {Yu}}, \bibinfo {author} {\bibfnamefont {X.}~\bibnamefont {Qiu}}, \bibinfo
  {author} {\bibfnamefont {Y.}~\bibnamefont {Wu}}, \bibinfo {author}
  {\bibfnamefont {J.}~\bibnamefont {Yoon}}, \bibinfo {author} {\bibfnamefont
  {P.}~\bibnamefont {Deorani}}, \bibinfo {author} {\bibfnamefont {J.~M.}\
  \bibnamefont {Besbas}}, \bibinfo {author} {\bibfnamefont {A.}~\bibnamefont
  {Manchon}},\ and\ \bibinfo {author} {\bibfnamefont {H.}~\bibnamefont
  {Yang}},\ }\bibfield  {title} {\bibinfo {title} {Spin orbit torques and
  Dzyaloshinskii-Moriya interaction in dual-interfaced Co-Ni multilayers},\
  }\href@noop {} {\bibfield  {journal} {\bibinfo  {journal} {Sci. Rep.}\
  }\textbf {\bibinfo {volume} {6}},\ \bibinfo {pages} {1} (\bibinfo {year}
  {2016})}\BibitemShut {NoStop}%
\bibitem [{\citenamefont {Soumyanarayanan}\ \emph {et~al.}(2017)\citenamefont
  {Soumyanarayanan}, \citenamefont {Raju}, \citenamefont {Oyarce},
  \citenamefont {Tan}, \citenamefont {Im}, \citenamefont {Petrovi{\'c}},
  \citenamefont {Ho}, \citenamefont {Khoo}, \citenamefont {Tran}, \citenamefont
  {Gan}, \citenamefont {Ernult},\ and\ \citenamefont
  {Panagopoulos}}]{soumyanarayanan2017tunable}%
  \BibitemOpen
  \bibfield  {author} {\bibinfo {author} {\bibfnamefont {A.}~\bibnamefont
  {Soumyanarayanan}}, \bibinfo {author} {\bibfnamefont {M.}~\bibnamefont
  {Raju}}, \bibinfo {author} {\bibfnamefont {A.~G.}\ \bibnamefont {Oyarce}},
  \bibinfo {author} {\bibfnamefont {A.~K.}\ \bibnamefont {Tan}}, \bibinfo
  {author} {\bibfnamefont {M.-Y.}\ \bibnamefont {Im}}, \bibinfo {author}
  {\bibfnamefont {A.~P.}\ \bibnamefont {Petrovi{\'c}}}, \bibinfo {author}
  {\bibfnamefont {P.}~\bibnamefont {Ho}}, \bibinfo {author} {\bibfnamefont
  {K.}~\bibnamefont {Khoo}}, \bibinfo {author} {\bibfnamefont {M.}~\bibnamefont
  {Tran}}, \bibinfo {author} {\bibfnamefont {C.}~\bibnamefont {Gan}}, \bibinfo
  {author} {\bibfnamefont {F.}~\bibnamefont {Ernult}},\ and\ \bibinfo {author}
  {\bibfnamefont {C.}~\bibnamefont {Panagopoulos}},\ }\bibfield  {title}
  {\bibinfo {title} {Tunable room-temperature magnetic skyrmions in Ir/Fe/Co/Pt
  multilayers},\ }\href@noop {} {\bibfield  {journal} {\bibinfo  {journal}
  {Nat. Mater.}\ }\textbf {\bibinfo {volume} {16}},\ \bibinfo {pages} {898}
  (\bibinfo {year} {2017})}\BibitemShut {NoStop}%
\bibitem [{\citenamefont {Kim}\ \emph {et~al.}(2017)\citenamefont {Kim},
  \citenamefont {Cho}, \citenamefont {Jung}, \citenamefont {Han}, \citenamefont
  {Yin}, \citenamefont {Kim}, \citenamefont {Swagten}, \citenamefont {Lee},
  \citenamefont {Jung},\ and\ \citenamefont {You}}]{kim2017}%
  \BibitemOpen
  \bibfield  {author} {\bibinfo {author} {\bibfnamefont {N.-H.}\ \bibnamefont
  {Kim}}, \bibinfo {author} {\bibfnamefont {J.}~\bibnamefont {Cho}}, \bibinfo
  {author} {\bibfnamefont {J.}~\bibnamefont {Jung}}, \bibinfo {author}
  {\bibfnamefont {D.-S.}\ \bibnamefont {Han}}, \bibinfo {author} {\bibfnamefont
  {Y.}~\bibnamefont {Yin}}, \bibinfo {author} {\bibfnamefont {J.-S.}\
  \bibnamefont {Kim}}, \bibinfo {author} {\bibfnamefont {H.~J.~M.}\
  \bibnamefont {Swagten}}, \bibinfo {author} {\bibfnamefont {K.}~\bibnamefont
  {Lee}}, \bibinfo {author} {\bibfnamefont {M.-H.}\ \bibnamefont {Jung}},\ and\
  \bibinfo {author} {\bibfnamefont {C.-Y.}\ \bibnamefont {You}},\ }\bibfield
  {title} {\bibinfo {title} {Role of top and bottom interfaces of a Pt/Co/AlOx
  system in Dzyaloshinskii-Moriya interaction, interface perpendicular magnetic
  anisotropy, and magneto-optical Kerr effect},\ }\href@noop {} {\bibfield
  {journal} {\bibinfo  {journal} {AIP Adv.}\ }\textbf {\bibinfo {volume} {7}},\
  \bibinfo {pages} {035213} (\bibinfo {year} {2017})}\BibitemShut {NoStop}%
\bibitem [{\citenamefont {Tolley}\ \emph {et~al.}(2018)\citenamefont {Tolley},
  \citenamefont {Montoya},\ and\ \citenamefont {Fullerton}}]{tolley2018}%
  \BibitemOpen
  \bibfield  {author} {\bibinfo {author} {\bibfnamefont {R.}~\bibnamefont
  {Tolley}}, \bibinfo {author} {\bibfnamefont {S.~A.}\ \bibnamefont
  {Montoya}},\ and\ \bibinfo {author} {\bibfnamefont {E.~E.}\ \bibnamefont
  {Fullerton}},\ }\bibfield  {title} {\bibinfo {title} {Room-temperature
  observation and current control of skyrmions in Pt/Co/Os/Pt thin films},\
  }\href@noop {} {\bibfield  {journal} {\bibinfo  {journal} {Phys. Rev.
  Mater.}\ }\textbf {\bibinfo {volume} {2}},\ \bibinfo {pages} {044404}
  (\bibinfo {year} {2018})}\BibitemShut {NoStop}%
\bibitem [{\citenamefont {Parakkat}\ \emph {et~al.}(2016)\citenamefont
  {Parakkat}, \citenamefont {Ganesh},\ and\ \citenamefont
  {Anil~Kumar}}]{parakkat2016}%
  \BibitemOpen
  \bibfield  {author} {\bibinfo {author} {\bibfnamefont {V.~M.}\ \bibnamefont
  {Parakkat}}, \bibinfo {author} {\bibfnamefont {K.~R.}\ \bibnamefont
  {Ganesh}},\ and\ \bibinfo {author} {\bibfnamefont {P.~S.}\ \bibnamefont
  {Anil~Kumar}},\ }\bibfield  {title} {\bibinfo {title} {Copper dusting effects
  on perpendicular magnetic anisotropy in Pt/Co/Pt tri-layers},\ }\href@noop {}
  {\bibfield  {journal} {\bibinfo  {journal} {AIP Adv.}\ }\textbf {\bibinfo
  {volume} {6}},\ \bibinfo {pages} {056122} (\bibinfo {year}
  {2016})}\BibitemShut {NoStop}%
\bibitem [{\citenamefont {Legrand}\ \emph {et~al.}(2018)\citenamefont
  {Legrand}, \citenamefont {Chauleau}, \citenamefont {Maccariello},
  \citenamefont {Reyren}, \citenamefont {Collin}, \citenamefont {Bouzehouane},
  \citenamefont {Jaouen}, \citenamefont {Cros},\ and\ \citenamefont
  {Fert}}]{Legrand2018}%
  \BibitemOpen
  \bibfield  {author} {\bibinfo {author} {\bibfnamefont {W.}~\bibnamefont
  {Legrand}}, \bibinfo {author} {\bibfnamefont {J.-Y.}\ \bibnamefont
  {Chauleau}}, \bibinfo {author} {\bibfnamefont {D.}~\bibnamefont
  {Maccariello}}, \bibinfo {author} {\bibfnamefont {N.}~\bibnamefont {Reyren}},
  \bibinfo {author} {\bibfnamefont {S.}~\bibnamefont {Collin}}, \bibinfo
  {author} {\bibfnamefont {K.}~\bibnamefont {Bouzehouane}}, \bibinfo {author}
  {\bibfnamefont {N.}~\bibnamefont {Jaouen}}, \bibinfo {author} {\bibfnamefont
  {V.}~\bibnamefont {Cros}},\ and\ \bibinfo {author} {\bibfnamefont
  {A.}~\bibnamefont {Fert}},\ }\bibfield  {title} {\bibinfo {title} {Hybrid
  chiral domain walls and skyrmions in magnetic multilayers},\ }\href@noop {}
  {\bibfield  {journal} {\bibinfo  {journal} {Sci. Adv.}\ }\textbf {\bibinfo
  {volume} {4}} (\bibinfo {year} {2018})}\BibitemShut {NoStop}%
\bibitem [{\citenamefont {Wells}\ \emph {et~al.}(2017)\citenamefont {Wells},
  \citenamefont {Shepley}, \citenamefont {Marrows},\ and\ \citenamefont
  {Moore}}]{wells2017}%
  \BibitemOpen
  \bibfield  {author} {\bibinfo {author} {\bibfnamefont {A.~W.~J.}\
  \bibnamefont {Wells}}, \bibinfo {author} {\bibfnamefont {P.~M.}\ \bibnamefont
  {Shepley}}, \bibinfo {author} {\bibfnamefont {C.~H.}\ \bibnamefont
  {Marrows}},\ and\ \bibinfo {author} {\bibfnamefont {T.~A.}\ \bibnamefont
  {Moore}},\ }\bibfield  {title} {\bibinfo {title} {Effect of interfacial
  intermixing on the Dzyaloshinskii-Moriya interaction in Pt/Co/Pt},\
  }\href@noop {} {\bibfield  {journal} {\bibinfo  {journal} {Phys. Rev. B}\
  }\textbf {\bibinfo {volume} {95}},\ \bibinfo {pages} {054428} (\bibinfo
  {year} {2017})}\BibitemShut {NoStop}%
\bibitem [{\citenamefont {Danilov}\ \emph {et~al.}(2018)\citenamefont
  {Danilov}, \citenamefont {Scherbakov}, \citenamefont {Glavin}, \citenamefont
  {Linnik}, \citenamefont {Kalashnikova}, \citenamefont {Shelukhin},
  \citenamefont {Pattnaik}, \citenamefont {Rushforth}, \citenamefont {Love},
  \citenamefont {Cavill}, \citenamefont {Yakovlev},\ and\ \citenamefont
  {Bayer}}]{danilov2018}%
  \BibitemOpen
  \bibfield  {author} {\bibinfo {author} {\bibfnamefont {A.~P.}\ \bibnamefont
  {Danilov}}, \bibinfo {author} {\bibfnamefont {A.~V.}\ \bibnamefont
  {Scherbakov}}, \bibinfo {author} {\bibfnamefont {B.~A.}\ \bibnamefont
  {Glavin}}, \bibinfo {author} {\bibfnamefont {T.~L.}\ \bibnamefont {Linnik}},
  \bibinfo {author} {\bibfnamefont {A.~M.}\ \bibnamefont {Kalashnikova}},
  \bibinfo {author} {\bibfnamefont {L.~A.}\ \bibnamefont {Shelukhin}}, \bibinfo
  {author} {\bibfnamefont {D.~P.}\ \bibnamefont {Pattnaik}}, \bibinfo {author}
  {\bibfnamefont {A.~W.}\ \bibnamefont {Rushforth}}, \bibinfo {author}
  {\bibfnamefont {C.~J.}\ \bibnamefont {Love}}, \bibinfo {author}
  {\bibfnamefont {S.~A.}\ \bibnamefont {Cavill}}, \bibinfo {author}
  {\bibfnamefont {D.~R.}\ \bibnamefont {Yakovlev}},\ and\ \bibinfo {author}
  {\bibfnamefont {M.}~\bibnamefont {Bayer}},\ }\bibfield  {title} {\bibinfo
  {title} {Optically excited spin pumping mediating collective magnetization
  dynamics in a spin valve structure},\ }\href@noop {} {\bibfield  {journal}
  {\bibinfo  {journal} {Phys. Rev. B}\ }\textbf {\bibinfo {volume} {98}},\
  \bibinfo {pages} {060406} (\bibinfo {year} {2018})}\BibitemShut {NoStop}%
\bibitem [{\citenamefont {Belmeguenai}\ \emph {et~al.}(2017)\citenamefont
  {Belmeguenai}, \citenamefont {Bouloussa}, \citenamefont {Roussign\'e},
  \citenamefont {Gabor}, \citenamefont {Petrisor}, \citenamefont {Tiusan},
  \citenamefont {Yang}, \citenamefont {Stashkevich},\ and\ \citenamefont
  {Ch\'erif}}]{belmeguenai2017}%
  \BibitemOpen
  \bibfield  {author} {\bibinfo {author} {\bibfnamefont {M.}~\bibnamefont
  {Belmeguenai}}, \bibinfo {author} {\bibfnamefont {H.}~\bibnamefont
  {Bouloussa}}, \bibinfo {author} {\bibfnamefont {Y.}~\bibnamefont
  {Roussign\'e}}, \bibinfo {author} {\bibfnamefont {M.~S.}\ \bibnamefont
  {Gabor}}, \bibinfo {author} {\bibfnamefont {T.}~\bibnamefont {Petrisor}},
  \bibinfo {author} {\bibfnamefont {C.}~\bibnamefont {Tiusan}}, \bibinfo
  {author} {\bibfnamefont {H.}~\bibnamefont {Yang}}, \bibinfo {author}
  {\bibfnamefont {A.}~\bibnamefont {Stashkevich}},\ and\ \bibinfo {author}
  {\bibfnamefont {S.~M.}\ \bibnamefont {Ch\'erif}},\ }\bibfield  {title}
  {\bibinfo {title} {Interface Dzyaloshinskii-Moriya interaction in the
  interlayer antiferromagnetic-exchange coupled Pt/CoFeB/Ru/CoFeB systems},\
  }\href@noop {} {\bibfield  {journal} {\bibinfo  {journal} {Phys. Rev. B}\
  }\textbf {\bibinfo {volume} {96}},\ \bibinfo {pages} {144402} (\bibinfo
  {year} {2017})}\BibitemShut {NoStop}%
\bibitem [{\citenamefont {dos Santos}\ \emph
  {et~al.}(2020{\natexlab{b}})\citenamefont {dos Santos}, \citenamefont {dos
  Santos~Dias},\ and\ \citenamefont {Lounis}}]{santos2020modelling}%
  \BibitemOpen
  \bibfield  {author} {\bibinfo {author} {\bibfnamefont {F.~J.}\ \bibnamefont
  {dos Santos}}, \bibinfo {author} {\bibfnamefont {M.}~\bibnamefont {dos
  Santos~Dias}},\ and\ \bibinfo {author} {\bibfnamefont {S.}~\bibnamefont
  {Lounis}},\ }\bibfield  {title} {\bibinfo {title} {Modeling spin waves in
  noncollinear antiferromagnets: Spin-flop states, spin spirals, skyrmions, and
  antiskyrmions},\ }\href {https://doi.org/10.1103/PhysRevB.102.104436}
  {\bibfield  {journal} {\bibinfo  {journal} {Phys. Rev. B}\ }\textbf {\bibinfo
  {volume} {102}},\ \bibinfo {pages} {104436} (\bibinfo {year}
  {2020}{\natexlab{b}})}\BibitemShut {NoStop}%
\bibitem [{\citenamefont {Costa}\ \emph {et~al.}(2010)\citenamefont {Costa},
  \citenamefont {Muniz}, \citenamefont {Lounis}, \citenamefont {Klautau},\ and\
  \citenamefont {Mills}}]{costa2010}%
  \BibitemOpen
  \bibfield  {author} {\bibinfo {author} {\bibfnamefont {A.~T.}\ \bibnamefont
  {Costa}}, \bibinfo {author} {\bibfnamefont {R.~B.}\ \bibnamefont {Muniz}},
  \bibinfo {author} {\bibfnamefont {S.}~\bibnamefont {Lounis}}, \bibinfo
  {author} {\bibfnamefont {A.~B.}\ \bibnamefont {Klautau}},\ and\ \bibinfo
  {author} {\bibfnamefont {D.~L.}\ \bibnamefont {Mills}},\ }\bibfield  {title}
  {\bibinfo {title} {Spin-orbit coupling and spin waves in ultrathin
  ferromagnets: The spin-wave Rashba effect},\ }\href@noop {} {\bibfield
  {journal} {\bibinfo  {journal} {Phys. Rev. B}\ }\textbf {\bibinfo {volume}
  {82}},\ \bibinfo {pages} {014428} (\bibinfo {year} {2010})}\BibitemShut
  {NoStop}%
\bibitem [{\citenamefont {Henry}\ \emph {et~al.}(2019)\citenamefont {Henry},
  \citenamefont {Stoeffler}, \citenamefont {Kim},\ and\ \citenamefont
  {Bailleul}}]{henry2019}%
  \BibitemOpen
  \bibfield  {author} {\bibinfo {author} {\bibfnamefont {Y.}~\bibnamefont
  {Henry}}, \bibinfo {author} {\bibfnamefont {D.}~\bibnamefont {Stoeffler}},
  \bibinfo {author} {\bibfnamefont {J.-V.}\ \bibnamefont {Kim}},\ and\ \bibinfo
  {author} {\bibfnamefont {M.}~\bibnamefont {Bailleul}},\ }\bibfield  {title}
  {\bibinfo {title} {Unidirectional spin-wave channeling along magnetic domain
  walls of Bloch type},\ }\href@noop {} {\bibfield  {journal} {\bibinfo
  {journal} {Phys. Rev. B}\ }\textbf {\bibinfo {volume} {100}},\ \bibinfo
  {pages} {024416} (\bibinfo {year} {2019})}\BibitemShut {NoStop}%
\bibitem [{\citenamefont {Grimsditch}\ \emph {et~al.}(1996)\citenamefont
  {Grimsditch}, \citenamefont {Kumar},\ and\ \citenamefont
  {Fullerton}}]{grimsditch1996}%
  \BibitemOpen
  \bibfield  {author} {\bibinfo {author} {\bibfnamefont {M.}~\bibnamefont
  {Grimsditch}}, \bibinfo {author} {\bibfnamefont {S.}~\bibnamefont {Kumar}},\
  and\ \bibinfo {author} {\bibfnamefont {E.~E.}\ \bibnamefont {Fullerton}},\
  }\bibfield  {title} {\bibinfo {title} {Brillouin light scattering study of
  Fe/Cr/Fe (211) and (100) trilayers},\ }\href@noop {} {\bibfield  {journal}
  {\bibinfo  {journal} {Phys. Rev. B}\ }\textbf {\bibinfo {volume} {54}},\
  \bibinfo {pages} {3385} (\bibinfo {year} {1996})}\BibitemShut {NoStop}%
\bibitem [{\citenamefont {Fuchs}\ \emph {et~al.}(1997)\citenamefont {Fuchs},
  \citenamefont {Ramsperger}, \citenamefont {Vaterlaus},\ and\ \citenamefont
  {Landolt}}]{fuchs1997}%
  \BibitemOpen
  \bibfield  {author} {\bibinfo {author} {\bibfnamefont {P.}~\bibnamefont
  {Fuchs}}, \bibinfo {author} {\bibfnamefont {U.}~\bibnamefont {Ramsperger}},
  \bibinfo {author} {\bibfnamefont {A.}~\bibnamefont {Vaterlaus}},\ and\
  \bibinfo {author} {\bibfnamefont {M.}~\bibnamefont {Landolt}},\ }\bibfield
  {title} {\bibinfo {title} {Roughness-induced coupling between ferromagnetic
  films across an amorphous spacer layer},\ }\href@noop {} {\bibfield
  {journal} {\bibinfo  {journal} {Phys. Rev. B}\ }\textbf {\bibinfo {volume}
  {55}},\ \bibinfo {pages} {12546} (\bibinfo {year} {1997})}\BibitemShut
  {NoStop}%
\bibitem [{\citenamefont {Fern{\'a}ndez-Pacheco}\ \emph
  {et~al.}(2019)\citenamefont {Fern{\'a}ndez-Pacheco}, \citenamefont
  {Vedmedenko}, \citenamefont {Ummelen}, \citenamefont {Mansell}, \citenamefont
  {Petit},\ and\ \citenamefont {Cowburn}}]{fernandez2019symmetry}%
  \BibitemOpen
  \bibfield  {author} {\bibinfo {author} {\bibfnamefont {A.}~\bibnamefont
  {Fern{\'a}ndez-Pacheco}}, \bibinfo {author} {\bibfnamefont {E.}~\bibnamefont
  {Vedmedenko}}, \bibinfo {author} {\bibfnamefont {F.}~\bibnamefont {Ummelen}},
  \bibinfo {author} {\bibfnamefont {R.}~\bibnamefont {Mansell}}, \bibinfo
  {author} {\bibfnamefont {D.}~\bibnamefont {Petit}},\ and\ \bibinfo {author}
  {\bibfnamefont {R.~P.}\ \bibnamefont {Cowburn}},\ }\bibfield  {title}
  {\bibinfo {title} {Symmetry-breaking interlayer Dzyaloshinskii--Moriya
  interactions in synthetic antiferromagnets},\ }\href@noop {} {\bibfield
  {journal} {\bibinfo  {journal} {Nat. Mater.}\ }\textbf {\bibinfo {volume}
  {18}},\ \bibinfo {pages} {679} (\bibinfo {year} {2019})}\BibitemShut
  {NoStop}%
\bibitem [{\citenamefont {Karnad}\ \emph {et~al.}(2018)\citenamefont {Karnad},
  \citenamefont {Freimuth}, \citenamefont {Martinez}, \citenamefont {Lo~Conte},
  \citenamefont {Gubbiotti}, \citenamefont {Schulz}, \citenamefont {Senz},
  \citenamefont {Ocker}, \citenamefont {Mokrousov},\ and\ \citenamefont
  {Kl\"aui}}]{karnad2018}%
  \BibitemOpen
  \bibfield  {author} {\bibinfo {author} {\bibfnamefont {G.~V.}\ \bibnamefont
  {Karnad}}, \bibinfo {author} {\bibfnamefont {F.}~\bibnamefont {Freimuth}},
  \bibinfo {author} {\bibfnamefont {E.}~\bibnamefont {Martinez}}, \bibinfo
  {author} {\bibfnamefont {R.}~\bibnamefont {Lo~Conte}}, \bibinfo {author}
  {\bibfnamefont {G.}~\bibnamefont {Gubbiotti}}, \bibinfo {author}
  {\bibfnamefont {T.}~\bibnamefont {Schulz}}, \bibinfo {author} {\bibfnamefont
  {S.}~\bibnamefont {Senz}}, \bibinfo {author} {\bibfnamefont {B.}~\bibnamefont
  {Ocker}}, \bibinfo {author} {\bibfnamefont {Y.}~\bibnamefont {Mokrousov}},\
  and\ \bibinfo {author} {\bibfnamefont {M.}~\bibnamefont {Kl\"aui}},\
  }\bibfield  {title} {\bibinfo {title} {Modification of
  Dzyaloshinskii-Moriya-interaction-stabilized domain wall chirality by driving
  currents},\ }\href@noop {} {\bibfield  {journal} {\bibinfo  {journal} {Phys.
  Rev. Lett.}\ }\textbf {\bibinfo {volume} {121}},\ \bibinfo {pages} {147203}
  (\bibinfo {year} {2018})}\BibitemShut {NoStop}%
\bibitem [{\citenamefont {Shahbazi}\ \emph {et~al.}(2019)\citenamefont
  {Shahbazi}, \citenamefont {Kim}, \citenamefont {Nembach}, \citenamefont
  {Shaw}, \citenamefont {Bischof}, \citenamefont {Rossell}, \citenamefont
  {Jeudy}, \citenamefont {Moore},\ and\ \citenamefont
  {Marrows}}]{shahbazi2019}%
  \BibitemOpen
  \bibfield  {author} {\bibinfo {author} {\bibfnamefont {K.}~\bibnamefont
  {Shahbazi}}, \bibinfo {author} {\bibfnamefont {J.-V.}\ \bibnamefont {Kim}},
  \bibinfo {author} {\bibfnamefont {H.~T.}\ \bibnamefont {Nembach}}, \bibinfo
  {author} {\bibfnamefont {J.~M.}\ \bibnamefont {Shaw}}, \bibinfo {author}
  {\bibfnamefont {A.}~\bibnamefont {Bischof}}, \bibinfo {author} {\bibfnamefont
  {M.~D.}\ \bibnamefont {Rossell}}, \bibinfo {author} {\bibfnamefont
  {V.}~\bibnamefont {Jeudy}}, \bibinfo {author} {\bibfnamefont {T.~A.}\
  \bibnamefont {Moore}},\ and\ \bibinfo {author} {\bibfnamefont {C.~H.}\
  \bibnamefont {Marrows}},\ }\bibfield  {title} {\bibinfo {title} {Domain-wall
  motion and interfacial Dzyaloshinskii-Moriya interactions in
  $\mathrm{Pt}/\mathrm{Co}/\mathrm{Ir}({t}_{\mathrm{Ir}})/\mathrm{Ta}$
  multilayers},\ }\href@noop {} {\bibfield  {journal} {\bibinfo  {journal}
  {Phys. Rev. B}\ }\textbf {\bibinfo {volume} {99}},\ \bibinfo {pages} {094409}
  (\bibinfo {year} {2019})}\BibitemShut {NoStop}%
\bibitem [{\citenamefont {Han}(2016)}]{han2016}%
  \BibitemOpen
  \bibfield  {author} {\bibinfo {author} {\bibfnamefont {W.}~\bibnamefont
  {Han}},\ }\bibfield  {title} {\bibinfo {title} {Perspectives for spintronics
  in 2D materials},\ }\href@noop {} {\bibfield  {journal} {\bibinfo  {journal}
  {APL Mater.}\ }\textbf {\bibinfo {volume} {4}},\ \bibinfo {pages} {032401}
  (\bibinfo {year} {2016})}\BibitemShut {NoStop}%
\bibitem [{\citenamefont {Samardak}\ \emph {et~al.}(2018)\citenamefont
  {Samardak}, \citenamefont {Kolesnikov}, \citenamefont {Stebliy},
  \citenamefont {Chebotkevich}, \citenamefont {Sadovnikov}, \citenamefont
  {Nikitov}, \citenamefont {Talapatra}, \citenamefont {Mohanty},\ and\
  \citenamefont {Ognev}}]{samardak2018}%
  \BibitemOpen
  \bibfield  {author} {\bibinfo {author} {\bibfnamefont {A.}~\bibnamefont
  {Samardak}}, \bibinfo {author} {\bibfnamefont {A.}~\bibnamefont
  {Kolesnikov}}, \bibinfo {author} {\bibfnamefont {M.}~\bibnamefont {Stebliy}},
  \bibinfo {author} {\bibfnamefont {L.}~\bibnamefont {Chebotkevich}}, \bibinfo
  {author} {\bibfnamefont {A.}~\bibnamefont {Sadovnikov}}, \bibinfo {author}
  {\bibfnamefont {S.}~\bibnamefont {Nikitov}}, \bibinfo {author} {\bibfnamefont
  {A.}~\bibnamefont {Talapatra}}, \bibinfo {author} {\bibfnamefont
  {J.}~\bibnamefont {Mohanty}},\ and\ \bibinfo {author} {\bibfnamefont
  {A.}~\bibnamefont {Ognev}},\ }\bibfield  {title} {\bibinfo {title} {Enhanced
  interfacial Dzyaloshinskii-Moriya interaction and isolated skyrmions in the
  inversion-symmetry-broken Ru/Co/W/Ru films},\ }\href@noop {} {\bibfield
  {journal} {\bibinfo  {journal} {Appl. Phys. Lett.}\ }\textbf {\bibinfo
  {volume} {112}},\ \bibinfo {pages} {192406} (\bibinfo {year}
  {2018})}\BibitemShut {NoStop}%
\bibitem [{\citenamefont {Kuepferling}\ \emph {et~al.}(2020)\citenamefont
  {Kuepferling}, \citenamefont {Casiraghi}, \citenamefont {Soares},
  \citenamefont {Durin}, \citenamefont {Garcia-Sanchez}, \citenamefont {Chen},
  \citenamefont {Back}, \citenamefont {Marrows}, \citenamefont {Tacchi},\ and\
  \citenamefont {Carlotti}}]{kuepferling2020measuring}%
  \BibitemOpen
  \bibfield  {author} {\bibinfo {author} {\bibfnamefont {M.}~\bibnamefont
  {Kuepferling}}, \bibinfo {author} {\bibfnamefont {A.}~\bibnamefont
  {Casiraghi}}, \bibinfo {author} {\bibfnamefont {G.}~\bibnamefont {Soares}},
  \bibinfo {author} {\bibfnamefont {G.}~\bibnamefont {Durin}}, \bibinfo
  {author} {\bibfnamefont {F.}~\bibnamefont {Garcia-Sanchez}}, \bibinfo
  {author} {\bibfnamefont {L.}~\bibnamefont {Chen}}, \bibinfo {author}
  {\bibfnamefont {C.~H.}\ \bibnamefont {Back}}, \bibinfo {author}
  {\bibfnamefont {C.~H.}\ \bibnamefont {Marrows}}, \bibinfo {author}
  {\bibfnamefont {S.}~\bibnamefont {Tacchi}},\ and\ \bibinfo {author}
  {\bibfnamefont {G.}~\bibnamefont {Carlotti}},\ }\href@noop {} {\bibinfo
  {title} {Measuring interfacial Dzyaloshinskii-Moriya interaction in ultra
  thin films}} (\bibinfo {year} {2020}),\ \Eprint
  {https://arxiv.org/abs/2009.11830} {arXiv:2009.11830} \BibitemShut {NoStop}%
\bibitem [{\citenamefont {Krams}\ \emph {et~al.}(1992)\citenamefont {Krams},
  \citenamefont {Lauks}, \citenamefont {Stamps}, \citenamefont {Hillebrands},\
  and\ \citenamefont {G\"untherodt}}]{krams1992}%
  \BibitemOpen
  \bibfield  {author} {\bibinfo {author} {\bibfnamefont {P.}~\bibnamefont
  {Krams}}, \bibinfo {author} {\bibfnamefont {F.}~\bibnamefont {Lauks}},
  \bibinfo {author} {\bibfnamefont {R.~L.}\ \bibnamefont {Stamps}}, \bibinfo
  {author} {\bibfnamefont {B.}~\bibnamefont {Hillebrands}},\ and\ \bibinfo
  {author} {\bibfnamefont {G.}~\bibnamefont {G\"untherodt}},\ }\bibfield
  {title} {\bibinfo {title} {Magnetic anisotropies of ultrathin Co(001) films
  on Cu(001)},\ }\href@noop {} {\bibfield  {journal} {\bibinfo  {journal}
  {Phys. Rev. Lett.}\ }\textbf {\bibinfo {volume} {69}},\ \bibinfo {pages}
  {3674} (\bibinfo {year} {1992})}\BibitemShut {NoStop}%
\bibitem [{\citenamefont {Madami}\ \emph {et~al.}(2004)\citenamefont {Madami},
  \citenamefont {Tacchi}, \citenamefont {Carlotti}, \citenamefont {Gubbiotti},\
  and\ \citenamefont {Stamps}}]{madami2004}%
  \BibitemOpen
  \bibfield  {author} {\bibinfo {author} {\bibfnamefont {M.}~\bibnamefont
  {Madami}}, \bibinfo {author} {\bibfnamefont {S.}~\bibnamefont {Tacchi}},
  \bibinfo {author} {\bibfnamefont {G.}~\bibnamefont {Carlotti}}, \bibinfo
  {author} {\bibfnamefont {G.}~\bibnamefont {Gubbiotti}},\ and\ \bibinfo
  {author} {\bibfnamefont {R.~L.}\ \bibnamefont {Stamps}},\ }\bibfield  {title}
  {\bibinfo {title} {In situ Brillouin scattering study of the thickness
  dependence of magnetic anisotropy in uncovered and Cu-covered Fe/GaAs(100)
  ultrathin films},\ }\href@noop {} {\bibfield  {journal} {\bibinfo  {journal}
  {Phys. Rev. B}\ }\textbf {\bibinfo {volume} {69}},\ \bibinfo {pages} {144408}
  (\bibinfo {year} {2004})}\BibitemShut {NoStop}%
\bibitem [{\citenamefont {Nembach}\ \emph {et~al.}(2015)\citenamefont
  {Nembach}, \citenamefont {Shaw}, \citenamefont {Weiler}, \citenamefont
  {Ju{\'e}},\ and\ \citenamefont {Silva}}]{nembach2015linear}%
  \BibitemOpen
  \bibfield  {author} {\bibinfo {author} {\bibfnamefont {H.~T.}\ \bibnamefont
  {Nembach}}, \bibinfo {author} {\bibfnamefont {J.~M.}\ \bibnamefont {Shaw}},
  \bibinfo {author} {\bibfnamefont {M.}~\bibnamefont {Weiler}}, \bibinfo
  {author} {\bibfnamefont {E.}~\bibnamefont {Ju{\'e}}},\ and\ \bibinfo {author}
  {\bibfnamefont {T.~J.}\ \bibnamefont {Silva}},\ }\bibfield  {title} {\bibinfo
  {title} {Linear relation between Heisenberg exchange and interfacial
  Dzyaloshinskii--Moriya interaction in metal films},\ }\href@noop {}
  {\bibfield  {journal} {\bibinfo  {journal} {Nat. Phys.}\ }\textbf {\bibinfo
  {volume} {11}},\ \bibinfo {pages} {825} (\bibinfo {year} {2015})}\BibitemShut
  {NoStop}%
\bibitem [{\citenamefont {Di}\ \emph {et~al.}(2015)\citenamefont {Di},
  \citenamefont {Zhang}, \citenamefont {Lim}, \citenamefont {Ng}, \citenamefont
  {Kuok}, \citenamefont {Yu}, \citenamefont {Yoon}, \citenamefont {Qiu},\ and\
  \citenamefont {Yang}}]{di2015}%
  \BibitemOpen
  \bibfield  {author} {\bibinfo {author} {\bibfnamefont {K.}~\bibnamefont
  {Di}}, \bibinfo {author} {\bibfnamefont {V.~L.}\ \bibnamefont {Zhang}},
  \bibinfo {author} {\bibfnamefont {H.~S.}\ \bibnamefont {Lim}}, \bibinfo
  {author} {\bibfnamefont {S.~C.}\ \bibnamefont {Ng}}, \bibinfo {author}
  {\bibfnamefont {M.~H.}\ \bibnamefont {Kuok}}, \bibinfo {author}
  {\bibfnamefont {J.}~\bibnamefont {Yu}}, \bibinfo {author} {\bibfnamefont
  {J.}~\bibnamefont {Yoon}}, \bibinfo {author} {\bibfnamefont {X.}~\bibnamefont
  {Qiu}},\ and\ \bibinfo {author} {\bibfnamefont {H.}~\bibnamefont {Yang}},\
  }\bibfield  {title} {\bibinfo {title} {Direct observation of the
  Dzyaloshinskii-Moriya interaction in a Pt/Co/Ni film},\ }\href@noop {}
  {\bibfield  {journal} {\bibinfo  {journal} {Phys. Rev. Lett.}\ }\textbf
  {\bibinfo {volume} {114}},\ \bibinfo {pages} {047201} (\bibinfo {year}
  {2015})}\BibitemShut {NoStop}%
\bibitem [{\citenamefont {Tacchi}\ \emph {et~al.}(2017)\citenamefont {Tacchi},
  \citenamefont {Troncoso}, \citenamefont {Ahlberg}, \citenamefont {Gubbiotti},
  \citenamefont {Madami}, \citenamefont {\AA{}kerman},\ and\ \citenamefont
  {Landeros}}]{tacchi2017}%
  \BibitemOpen
  \bibfield  {author} {\bibinfo {author} {\bibfnamefont {S.}~\bibnamefont
  {Tacchi}}, \bibinfo {author} {\bibfnamefont {R.~E.}\ \bibnamefont
  {Troncoso}}, \bibinfo {author} {\bibfnamefont {M.}~\bibnamefont {Ahlberg}},
  \bibinfo {author} {\bibfnamefont {G.}~\bibnamefont {Gubbiotti}}, \bibinfo
  {author} {\bibfnamefont {M.}~\bibnamefont {Madami}}, \bibinfo {author}
  {\bibfnamefont {J.}~\bibnamefont {\AA{}kerman}},\ and\ \bibinfo {author}
  {\bibfnamefont {P.}~\bibnamefont {Landeros}},\ }\bibfield  {title} {\bibinfo
  {title} {Interfacial Dzyaloshinskii-Moriya interaction in
  $\mathrm{Pt}/\mathrm{CoFeB}$ films: Effect of the heavy-metal thickness},\
  }\href@noop {} {\bibfield  {journal} {\bibinfo  {journal} {Phys. Rev. Lett.}\
  }\textbf {\bibinfo {volume} {118}},\ \bibinfo {pages} {147201} (\bibinfo
  {year} {2017})}\BibitemShut {NoStop}%
\bibitem [{\citenamefont {Kumar}\ \emph {et~al.}(2020)\citenamefont {Kumar},
  \citenamefont {Chaurasiya}, \citenamefont {Chowdhury}, \citenamefont
  {Mondal}, \citenamefont {Bansal}, \citenamefont {Barvat}, \citenamefont
  {Khanna}, \citenamefont {Pal}, \citenamefont {Chaudhary}, \citenamefont
  {Barman},\ and\ \citenamefont {Muduli}}]{kumar2020}%
  \BibitemOpen
  \bibfield  {author} {\bibinfo {author} {\bibfnamefont {A.}~\bibnamefont
  {Kumar}}, \bibinfo {author} {\bibfnamefont {A.~K.}\ \bibnamefont
  {Chaurasiya}}, \bibinfo {author} {\bibfnamefont {N.}~\bibnamefont
  {Chowdhury}}, \bibinfo {author} {\bibfnamefont {A.~K.}\ \bibnamefont
  {Mondal}}, \bibinfo {author} {\bibfnamefont {R.}~\bibnamefont {Bansal}},
  \bibinfo {author} {\bibfnamefont {A.}~\bibnamefont {Barvat}}, \bibinfo
  {author} {\bibfnamefont {S.~P.}\ \bibnamefont {Khanna}}, \bibinfo {author}
  {\bibfnamefont {P.}~\bibnamefont {Pal}}, \bibinfo {author} {\bibfnamefont
  {S.}~\bibnamefont {Chaudhary}}, \bibinfo {author} {\bibfnamefont
  {A.}~\bibnamefont {Barman}},\ and\ \bibinfo {author} {\bibfnamefont {P.~K.}\
  \bibnamefont {Muduli}},\ }\bibfield  {title} {\bibinfo {title} {Direct
  measurement of interfacial Dzyaloshinskii–Moriya interaction at the
  MoS2/Ni80Fe20 interface},\ }\href@noop {} {\bibfield  {journal} {\bibinfo
  {journal} {Appl. Phys. Lett.}\ }\textbf {\bibinfo {volume} {116}},\ \bibinfo
  {pages} {232405} (\bibinfo {year} {2020})}\BibitemShut {NoStop}%
\bibitem [{\citenamefont {Bouloussa}\ \emph {et~al.}(2018)\citenamefont
  {Bouloussa}, \citenamefont {Roussign\'e}, \citenamefont {Belmeguenai},
  \citenamefont {Stashkevich}, \citenamefont {Ch\'erif}, \citenamefont {Yu},\
  and\ \citenamefont {Yang}}]{bouloussa2018}%
  \BibitemOpen
  \bibfield  {author} {\bibinfo {author} {\bibfnamefont {H.}~\bibnamefont
  {Bouloussa}}, \bibinfo {author} {\bibfnamefont {Y.}~\bibnamefont
  {Roussign\'e}}, \bibinfo {author} {\bibfnamefont {M.}~\bibnamefont
  {Belmeguenai}}, \bibinfo {author} {\bibfnamefont {A.}~\bibnamefont
  {Stashkevich}}, \bibinfo {author} {\bibfnamefont {S.~M.}\ \bibnamefont
  {Ch\'erif}}, \bibinfo {author} {\bibfnamefont {J.}~\bibnamefont {Yu}},\ and\
  \bibinfo {author} {\bibfnamefont {H.}~\bibnamefont {Yang}},\ }\bibfield
  {title} {\bibinfo {title} {Spin-wave calculations for magnetic stacks with
  interface Dzyaloshinskii-Moriya interaction},\ }\href@noop {} {\bibfield
  {journal} {\bibinfo  {journal} {Phys. Rev. B}\ }\textbf {\bibinfo {volume}
  {98}},\ \bibinfo {pages} {024428} (\bibinfo {year} {2018})}\BibitemShut
  {NoStop}%
\bibitem [{\citenamefont {Sandercock}(1982)}]{sandercock1982}%
  \BibitemOpen
  \bibfield  {author} {\bibinfo {author} {\bibfnamefont {J.~R.}\ \bibnamefont
  {Sandercock}},\ }\bibinfo {title} {Light scattering in solids III}\ (\bibinfo
   {publisher} {Springer-Verlag},\ \bibinfo {year} {1982})\ p.\ \bibinfo
  {pages} {173}\BibitemShut {NoStop}%
\bibitem [{\citenamefont {Carlotti}\ and\ \citenamefont
  {Gubbiotti}(2002)}]{Carlotti_2002}%
  \BibitemOpen
  \bibfield  {author} {\bibinfo {author} {\bibfnamefont {G.}~\bibnamefont
  {Carlotti}}\ and\ \bibinfo {author} {\bibfnamefont {G.}~\bibnamefont
  {Gubbiotti}},\ }\bibfield  {title} {\bibinfo {title} {Magnetic properties of
  layered nanostructures studied by means of Brillouin light scattering and the
  surface magneto-optical Kerr effect},\ }\href@noop {} {\bibfield  {journal}
  {\bibinfo  {journal} {J. Phys. Condens. Matter}\ }\textbf {\bibinfo {volume}
  {14}},\ \bibinfo {pages} {8199} (\bibinfo {year} {2002})}\BibitemShut
  {NoStop}%
\bibitem [{\citenamefont {Graczyk}\ \emph {et~al.}(2018)\citenamefont
  {Graczyk}, \citenamefont {Zelent},\ and\ \citenamefont
  {Krawczyk}}]{graczyk2018}%
  \BibitemOpen
  \bibfield  {author} {\bibinfo {author} {\bibfnamefont {P.}~\bibnamefont
  {Graczyk}}, \bibinfo {author} {\bibfnamefont {M.}~\bibnamefont {Zelent}},\
  and\ \bibinfo {author} {\bibfnamefont {M.}~\bibnamefont {Krawczyk}},\
  }\bibfield  {title} {\bibinfo {title} {{Co- and contra-directional vertical
  coupling between ferromagnetic layers with grating for short-wavelength spin
  wave generation}},\ }\href@noop {} {\bibfield  {journal} {\bibinfo  {journal}
  {New J. Phys.}\ }\textbf {\bibinfo {volume} {20}},\ \bibinfo {pages} {053021}
  (\bibinfo {year} {2018})}\BibitemShut {NoStop}%
\bibitem [{\citenamefont {Klingler}\ \emph {et~al.}(2018)\citenamefont
  {Klingler}, \citenamefont {Amin}, \citenamefont {Gepr\"ags}, \citenamefont
  {Ganzhorn}, \citenamefont {Maier-Flaig}, \citenamefont {Althammer},
  \citenamefont {Huebl}, \citenamefont {Gross}, \citenamefont {McMichael},
  \citenamefont {Stiles}, \citenamefont {Goennenwein},\ and\ \citenamefont
  {Weiler}}]{klingler2018}%
  \BibitemOpen
  \bibfield  {author} {\bibinfo {author} {\bibfnamefont {S.}~\bibnamefont
  {Klingler}}, \bibinfo {author} {\bibfnamefont {V.}~\bibnamefont {Amin}},
  \bibinfo {author} {\bibfnamefont {S.}~\bibnamefont {Gepr\"ags}}, \bibinfo
  {author} {\bibfnamefont {K.}~\bibnamefont {Ganzhorn}}, \bibinfo {author}
  {\bibfnamefont {H.}~\bibnamefont {Maier-Flaig}}, \bibinfo {author}
  {\bibfnamefont {M.}~\bibnamefont {Althammer}}, \bibinfo {author}
  {\bibfnamefont {H.}~\bibnamefont {Huebl}}, \bibinfo {author} {\bibfnamefont
  {R.}~\bibnamefont {Gross}}, \bibinfo {author} {\bibfnamefont {R.~D.}\
  \bibnamefont {McMichael}}, \bibinfo {author} {\bibfnamefont {M.~D.}\
  \bibnamefont {Stiles}}, \bibinfo {author} {\bibfnamefont {S.~T.~B.}\
  \bibnamefont {Goennenwein}},\ and\ \bibinfo {author} {\bibfnamefont
  {M.}~\bibnamefont {Weiler}},\ }\bibfield  {title} {\bibinfo {title}
  {Spin-torque excitation of perpendicular standing spin waves in coupled
  $\mathrm{YIG}/\mathrm{Co}$ heterostructures},\ }\href@noop {} {\bibfield
  {journal} {\bibinfo  {journal} {Phys. Rev. Lett.}\ }\textbf {\bibinfo
  {volume} {120}},\ \bibinfo {pages} {127201} (\bibinfo {year}
  {2018})}\BibitemShut {NoStop}%
\bibitem [{\citenamefont {Gubbiotti}\ \emph {et~al.}(1997)\citenamefont
  {Gubbiotti}, \citenamefont {Carlotti}, \citenamefont {Socino}, \citenamefont
  {D'Orazio}, \citenamefont {Lucari}, \citenamefont {Bernardini},\ and\
  \citenamefont {De~Crescenzi}}]{gubbiotti1997}%
  \BibitemOpen
  \bibfield  {author} {\bibinfo {author} {\bibfnamefont {G.}~\bibnamefont
  {Gubbiotti}}, \bibinfo {author} {\bibfnamefont {G.}~\bibnamefont {Carlotti}},
  \bibinfo {author} {\bibfnamefont {G.}~\bibnamefont {Socino}}, \bibinfo
  {author} {\bibfnamefont {F.}~\bibnamefont {D'Orazio}}, \bibinfo {author}
  {\bibfnamefont {F.}~\bibnamefont {Lucari}}, \bibinfo {author} {\bibfnamefont
  {R.}~\bibnamefont {Bernardini}},\ and\ \bibinfo {author} {\bibfnamefont
  {M.}~\bibnamefont {De~Crescenzi}},\ }\bibfield  {title} {\bibinfo {title}
  {Perpendicular and in-plane magnetic anisotropy in epitaxial Cu/Ni/Cu/Si(111)
  ultrathin films},\ }\href@noop {} {\bibfield  {journal} {\bibinfo  {journal}
  {Phys. Rev. B}\ }\textbf {\bibinfo {volume} {56}},\ \bibinfo {pages} {11073}
  (\bibinfo {year} {1997})}\BibitemShut {NoStop}%
\bibitem [{\citenamefont {Moon}\ \emph {et~al.}(2013)\citenamefont {Moon},
  \citenamefont {Seo}, \citenamefont {Lee}, \citenamefont {Kim}, \citenamefont
  {Ryu}, \citenamefont {Lee}, \citenamefont {McMichael},\ and\ \citenamefont
  {Stiles}}]{moon2013}%
  \BibitemOpen
  \bibfield  {author} {\bibinfo {author} {\bibfnamefont {J.-H.}\ \bibnamefont
  {Moon}}, \bibinfo {author} {\bibfnamefont {S.-M.}\ \bibnamefont {Seo}},
  \bibinfo {author} {\bibfnamefont {K.-J.}\ \bibnamefont {Lee}}, \bibinfo
  {author} {\bibfnamefont {K.-W.}\ \bibnamefont {Kim}}, \bibinfo {author}
  {\bibfnamefont {J.}~\bibnamefont {Ryu}}, \bibinfo {author} {\bibfnamefont
  {H.-W.}\ \bibnamefont {Lee}}, \bibinfo {author} {\bibfnamefont {R.~D.}\
  \bibnamefont {McMichael}},\ and\ \bibinfo {author} {\bibfnamefont {M.~D.}\
  \bibnamefont {Stiles}},\ }\bibfield  {title} {\bibinfo {title} {Spin-wave
  propagation in the presence of interfacial Dzyaloshinskii-Moriya
  interaction},\ }\href@noop {} {\bibfield  {journal} {\bibinfo  {journal}
  {Phys. Rev. B}\ }\textbf {\bibinfo {volume} {88}},\ \bibinfo {pages} {184404}
  (\bibinfo {year} {2013})}\BibitemShut {NoStop}%
\bibitem [{\citenamefont {Mendisch}\ \emph {et~al.}(2019)\citenamefont
  {Mendisch}, \citenamefont {Žiemys}, \citenamefont {Ahrens}, \citenamefont
  {Papp},\ and\ \citenamefont {Becherer}}]{MENDISCH2019345}%
  \BibitemOpen
  \bibfield  {author} {\bibinfo {author} {\bibfnamefont {S.}~\bibnamefont
  {Mendisch}}, \bibinfo {author} {\bibfnamefont {G.}~\bibnamefont {Žiemys}},
  \bibinfo {author} {\bibfnamefont {V.}~\bibnamefont {Ahrens}}, \bibinfo
  {author} {\bibfnamefont {A.}~\bibnamefont {Papp}},\ and\ \bibinfo {author}
  {\bibfnamefont {M.}~\bibnamefont {Becherer}},\ }\bibfield  {title} {\bibinfo
  {title} {Pt\textbackslash Co\textbackslash W as a candidate for low power
  nanomagnetic logic},\ }\href@noop {} {\bibfield  {journal} {\bibinfo
  {journal} {J. Magn. Magn. Mater.}\ }\textbf {\bibinfo {volume} {485}},\
  \bibinfo {pages} {345 } (\bibinfo {year} {2019})}\BibitemShut {NoStop}%
\bibitem [{\citenamefont {Chowdhury}\ \emph {et~al.}(2012)\citenamefont
  {Chowdhury}, \citenamefont {Kulkarni}, \citenamefont {Krishnan},
  \citenamefont {Barshilia}, \citenamefont {Sagdeo}, \citenamefont {Rai},
  \citenamefont {Lodha},\ and\ \citenamefont {Sridhara~Rao}}]{chowdhury2012}%
  \BibitemOpen
  \bibfield  {author} {\bibinfo {author} {\bibfnamefont {P.}~\bibnamefont
  {Chowdhury}}, \bibinfo {author} {\bibfnamefont {P.~D.}\ \bibnamefont
  {Kulkarni}}, \bibinfo {author} {\bibfnamefont {M.}~\bibnamefont {Krishnan}},
  \bibinfo {author} {\bibfnamefont {H.~C.}\ \bibnamefont {Barshilia}}, \bibinfo
  {author} {\bibfnamefont {A.}~\bibnamefont {Sagdeo}}, \bibinfo {author}
  {\bibfnamefont {S.~K.}\ \bibnamefont {Rai}}, \bibinfo {author} {\bibfnamefont
  {G.~S.}\ \bibnamefont {Lodha}},\ and\ \bibinfo {author} {\bibfnamefont
  {D.~V.}\ \bibnamefont {Sridhara~Rao}},\ }\bibfield  {title} {\bibinfo {title}
  {Effect of coherent to incoherent structural transition on magnetic
  anisotropy in Co/Pt multilayers},\ }\href@noop {} {\bibfield  {journal}
  {\bibinfo  {journal} {J. Appl. Phys.}\ }\textbf {\bibinfo {volume} {112}},\
  \bibinfo {pages} {023912} (\bibinfo {year} {2012})}\BibitemShut {NoStop}%
\bibitem [{\citenamefont {Mallick}\ \emph {et~al.}(2018)\citenamefont
  {Mallick}, \citenamefont {Mallik}, \citenamefont {Singh}, \citenamefont
  {Chowdhury}, \citenamefont {Gieniusz}, \citenamefont {Maziewski},\ and\
  \citenamefont {Bedanta}}]{Mallick_2018}%
  \BibitemOpen
  \bibfield  {author} {\bibinfo {author} {\bibfnamefont {S.}~\bibnamefont
  {Mallick}}, \bibinfo {author} {\bibfnamefont {S.}~\bibnamefont {Mallik}},
  \bibinfo {author} {\bibfnamefont {B.~B.}\ \bibnamefont {Singh}}, \bibinfo
  {author} {\bibfnamefont {N.}~\bibnamefont {Chowdhury}}, \bibinfo {author}
  {\bibfnamefont {R.}~\bibnamefont {Gieniusz}}, \bibinfo {author}
  {\bibfnamefont {A.}~\bibnamefont {Maziewski}},\ and\ \bibinfo {author}
  {\bibfnamefont {S.}~\bibnamefont {Bedanta}},\ }\bibfield  {title} {\bibinfo
  {title} {Tuning the anisotropy and domain structure of Co films by variable
  growth conditions and seed layers},\ }\href
  {https://doi.org/10.1088/1361-6463/aac880} {\bibfield  {journal} {\bibinfo
  {journal} {J. Phys. D: Appl. Phys.}\ }\textbf {\bibinfo {volume} {51}},\
  \bibinfo {pages} {275003} (\bibinfo {year} {2018})}\BibitemShut {NoStop}%
\end{thebibliography}%

\end{document}